\documentclass[pra,twocolumn,superscriptaddress]{revtex4-2}

\usepackage{amsmath}
\usepackage{amssymb}
\usepackage{graphicx}
\usepackage{upgreek}
\usepackage[colorlinks=true, citecolor=blue, urlcolor=blue]{hyperref}

\begin{document}
\title{Experimental quantification of  quantum coherence for a set of quantum states}
\author{Tianle Zheng}
\affiliation{School of Physical Science and Technology, Ningbo University, Ningbo, 315211, China}
\author{Liangsheng Li}
\affiliation{National Key Laboratory of Scattering and Radiation, Beijing 100854, China}
\author{Wenting Zhou}
\affiliation{School of Physical Science and Technology, Ningbo University, Ningbo, 315211, China}
\author{Chengjie Zhang}
\email{chengjie.zhang@gmail.com}
\affiliation{School of Physical Science and Technology, Ningbo University, Ningbo, 315211, China}

\begin{abstract}
We present a direct experimental verification of quantum coherence for a set of quantum states in a basis-independent manner. We find that the theory of quantum coherence for a set of quantum states matches our experimental results perfectly and can be applied to protocols like BB84 and quantum secure direct communication. 
Using a Sagnac interferometer,  we experimentally quantify the quantum coherence for two sets of experimental states. Moreover, we introduce a novel application of set coherence, illustrating that the states employed in the BB84 protocol exhibit maximal set coherence. Our results pave the way for wider applications of quantum coherence for a set of quantum states, including quantum key distribution protocols and probabilistic quantum cloning.
\end{abstract}

\maketitle
	
\section{Introduction}
The quantum state superposition is one of the fundamental principles in quantum mechanics. Quantum coherence, a critical resource for quantum information processing, arises directly from this state superposition. It is extensively utilized in various applications, such as quantum computing~\cite{Hillery16,Shi17},  quantum phase transitions~\cite{Malvezzi16,Li20,Pires21}, and quantum metrology~\cite{Castellini19,Pires18,McCormick19}. Moreover, the coherence of quantum states distinguishes quantum physics from classical physics, underpinning the inherent randomness of quantum mechanics and forming the basis for the phenomenon of entanglement.

Numerous methods for measuring coherence have been explored in various works, including distance-based coherence measures~\cite{ Baumgratz14,Chen16,Rana16,Yao16062339,Hu17,Streltsov17}, entanglement-based coherence measures~\cite{Streltsov15}, convex roof measures of coherence~\cite{Yuan15, Qi17, Zhao18, Li19}, robustness of coherence~\cite{Napoli16, Piani16, Wang17}, skew-information-based measures of coherence~\cite{Girolami14, Du1536}, and other coherence measures~\cite{Yao16, Yu16, Vicente17}.
However, previous studies are often constrained by the choice of basis, and coherence measures are typically proposed for individual quantum states. To address this limitation, Ref.~\cite{Designolle21} introduced the innovative concept of set coherence, which characterizes the coherence of a set of quantum states without any preferred basis. This perspective is particularly compelling. The key insight of their work is the shift from focusing solely on the coherence of individual quantum states to considering the coherence of a set of states, termed set coherence. They suggest that a meaningful measure can be derived by minimizing the discrepancies among all possible basis choices for the states in a set. This method enables the quantitative analysis of set coherence without the constraints of any preferred basis.

Recent advancements in coherence theory have led to significant experimental breakthroughs.  Researchers have focused on various aspects related to coherence, aiming to deepen our understanding and harness its potential. For instance, coherence and state conversion~\cite{ Wu20}, distillation of quantum coherence~\cite{K.-D. Wu17, T. Zhang24,  T. Zhang24, H.Q.C. Wang22}, non-Fock state~\cite{G. Park23}, coherence and quantum correlations~\cite{ K.D. Wu18, B.S. Ham23, B.S. Ham24}, and others on coherence~\cite{C. Zhang19, Gianani23, G.W. Qian23, J. Lin24, W. Huang23}
are areas of great interest today.
Quantum state tomography stands as the prevalent technique for exploring coherence theory, necessitating multiple measurements from diverse perspectives to gather data about the quantum system. Alternative approaches~\cite{Wang17} exist that can directly measure the degree of quantum coherence by using the interference fringes. Coherence witnesses~\cite{Giordani23}, which do not require complete state tomography, offer a more economical solution for larger qubit systems. However, most experiments measure the coherence only for a single quantum state, and experimental results quantifying quantum coherence for an ensemble of quantum states remain very limited.

In this work, 
we experimentally investigate the basis-independent set coherence. By employing a Sagnac interferometer, we evaluate the quantum coherence of two separate sets of states. The findings of our study affirm that this technique is capable of providing a precise quantitative analysis of set coherence for quantum states in an experimental setting, independent of the chosen basis. Furthermore, we propose an application of set coherence, by proving that the states used in the BB84 protocol  are just ones with maximal set coherence.

\section{Theoretical analysis}
As mentioned in the introduction, usually one may choose the basis in the study of quantum coherence for  individual  quantum states, as most coherence measures are relative to the selected basis. However, defining and quantifying the coherence without a specific basis is also important.  Ref.~\cite{Designolle21} proposed a basis-independent method for quantifying set coherence, which has been defined by considering the coherence of a set of quantum states and quantified it by minimizing the coherence of this set under all possible basis choices. This measure allows one to identify the optimal basis that minimizes set coherence.

Consider a  set of $n$ quantum states  $\{\rho_{j}\}_{j=1}^n$ in a Hilbert space of dimension $d$,
Ref.~\cite{Designolle21} defined max robustness of set coherence as
$
  \mathcal{R}(\vec{\rho})=\min_{U}\max_{j}\mathcal{R}_{F_{1}}(U\rho_{j}U^{\dagger}),
$
where the unitary operator $U$ represents the choice of basis, $\mathcal{R}_{F_1}$ is the robustness of coherence for each individual state,  and $\vec{\rho}$ represents the entire set of $n$ quantum states.
However, directly calculating the maximum robustness of set coherence is a challenging task. Therefore, another method is proposed, i.e., replacing the maximum value with the average value of all states. The mean robustness of set coherence is obtained as \cite{Designolle21},
\begin{equation}\label{eq1}
  \mathcal{R}_{1}(\vec{\rho})=\min_U\frac{1}{n}\sum_{j=1}^{n}\mathcal{R}_{F_1} (U\rho_j U^\dag).
\end{equation}
For single qubit systems, it can be reduced to $\mathcal{R}_{1}(\vec{\rho})= \min_{\vec{p}\in S^{2}} \frac{1}{n}\sum_{j=1}^{n}\|\vec{q}_{j}\||\sin{(\vec{p},\vec{q}_{j})}|$~\cite{Designolle21},
where $(\vec{p},\vec{q}_{j})$ represents the angle between vectors $\vec{p}$ and $\vec{q}_{j}$,  the minimization is performed over all possible unit-length vectors $\vec{p}$ on the Bloch sphere $S^2$,
$\vec{q}_{j}$ resprents the Bloch vector of the state $\rho_j$, and $n$ is the number of states in set $\vec{\rho}$. Ref.~\cite{Designolle21} also provides specific examples, such as for pairs of pure qubit states. These methods of selecting optimal basis can improve the accuracy and meaningfulness of quantifying quantum coherence.

\begin{figure}[t]
\begin{center}
\includegraphics[scale=0.6]{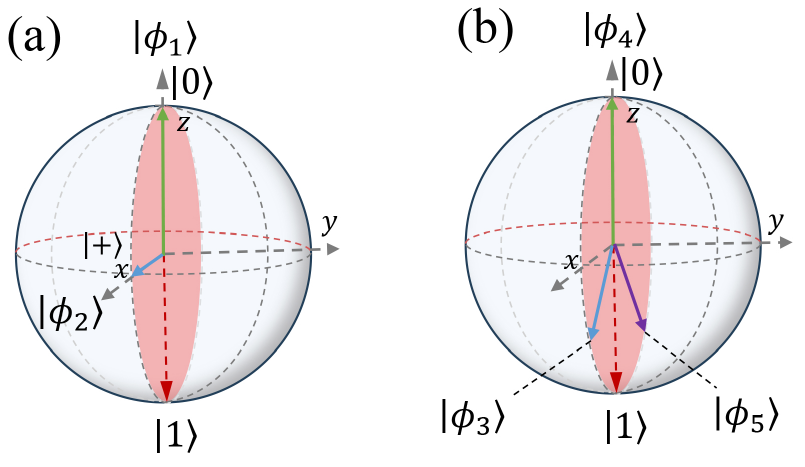}
\caption{(a) The Bloch sphere illustrates two pure states corresponding to $n=2$, which are mapped onto the X-axis and Z-axis of the Bloch sphere as $|\phi_{1}\rangle$ and $|\phi_{2}\rangle$, respectively. (b) The Bloch sphere represents  three pure states corresponding to $n=3$, which are denoted by $|\phi_{3}\rangle$, $|\phi_{4}\rangle$, and $|\phi_{5}\rangle$.} \label{fig1}
\end{center}
\end{figure}

In our experiments, we select two different sets of pure qubit states to experimentally investigate the set coherence. The first set is $\vec{\varrho_1}=\{|\phi_{1}\rangle,|\phi_{2}\rangle\}$ with $n=2$, where
$|\phi_{1}\rangle=|0\rangle$ and $|\phi_{2}\rangle=|+\rangle=(|0\rangle+|1\rangle)/\sqrt{2}$.
For pairs of pure qubit states, the minimum vector $\vec{p}$
in $\mathcal{R}_{1}(\vec{\rho})= \min_{\vec{p}\in S^{2}} \frac{1}{2}\sum_{j=1}^{2}\|\vec{q}_{j}\||\sin{(\vec{p},\vec{q}_{j})}|$ is either $\vec{q}_1$ or $\vec{q}_2$, i.e., $\mathcal{R}_{1}(\vec{\rho})=\frac{1}{2}|\sin(\vec{q}_1,\vec{q}_2)|\leq\frac{1}{2}$~\cite{Designolle21}.~The second set is $\vec{\varrho_2}=\{|\phi_{3}\rangle,|\phi_{4}\rangle,|\phi_{5}\rangle\}$
with $n=3$, where
$
|\phi_{3}\rangle=(|0\rangle+\sqrt{2}|1\rangle)/\sqrt{3},~
|\phi_{4}\rangle=|0\rangle,~
|\phi_{5}\rangle=(|0\rangle-\sqrt{2}|1\rangle)/\sqrt{3}.
$
When  $n=3$, the largest set coherence $\mathcal{R}_{1}(\vec{\rho})$ is attained  when the three Bloch vectors
form an orthonormal basis, and  $\mathcal{R}_{1}(\vec{\rho})$   reaches its upper bound $\mathcal{R}_{1}^{*}=2/3$~\cite{Designolle21}. Thus, one can conclude that the set coherence of $\{|\phi_{3}\rangle,|\phi_{4}\rangle,|\phi_{5}\rangle\}$ must not be greater than $2/3$. The  Bloch sphere representation of the states we selected in the cases of $n=2$ and $n=3$ is shown in Fig. \ref{fig1}.

\begin{figure*}[t]
\begin{center}
\includegraphics[scale=0.6]{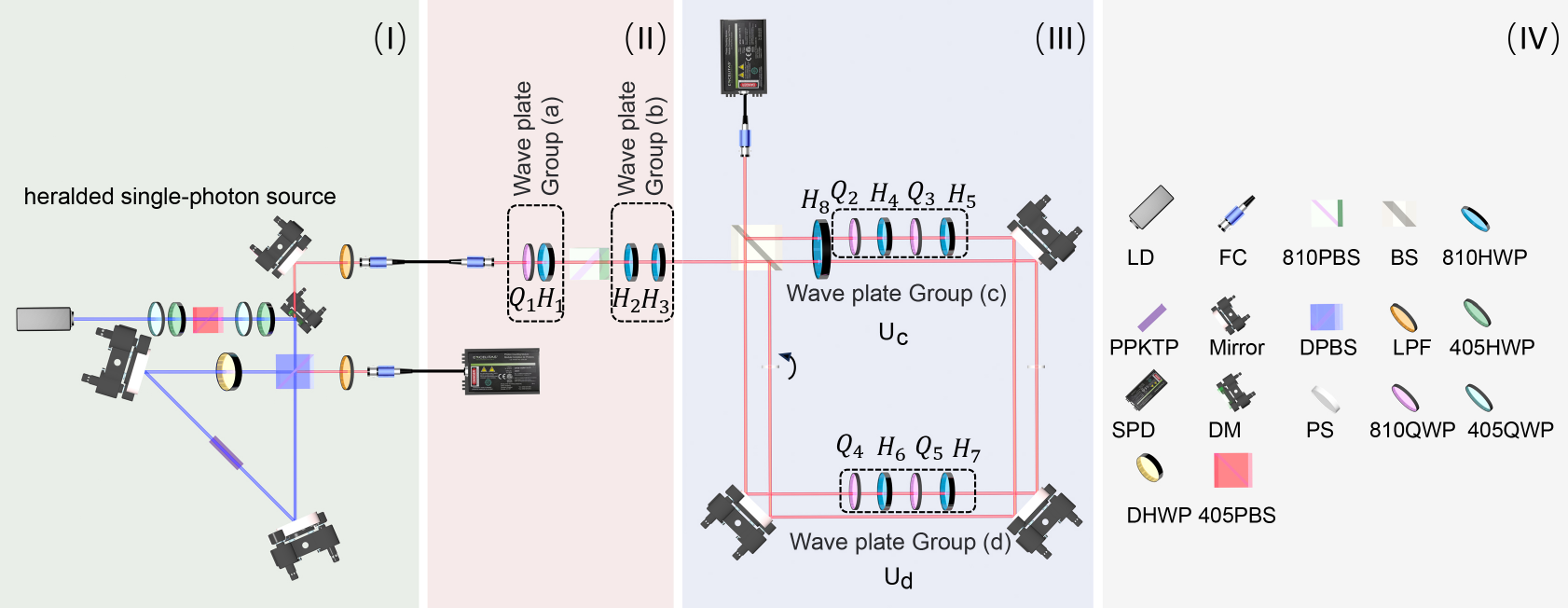}
\caption{The diagram of the experimental apparatus consists of four parts. (I) The first part is the heralded single-photon source, which utilizes the down-conversion effect of PPKTP crystal parameters to convert one high-energy photon into two lower-energy photons, thus generating entangled photons. By observing the change in the quantity of entangled photon pairs, interference visibility can be reflected. (II) The second part is about the preparation of the initial state. By using a wave plate group (a) and a PBS crystal, the intensity of photons in the third part can be adjusted to match the number of photon pairs. $\text{H}_{2}$ can convert horizontally polarized photons into arbitrarily polarized photons at any angle to simulate superposition states, while $\text{H}_{3}$ can change the measurement basis. (III) The third part is the main experimental section, where the number of HWPs and QWPs in wave plate groups (c) and (d) are equal to ensure that both paths in the interference ring have the same optical path length. The frontmost half-wave plate is used to counteract the polarization effect caused by the reflection mirror. (IV) In the fourth section, we listed all the optical devices used in this experiment along with their abbreviations.} \label{fig2}
\end{center}
\end{figure*}

Now we discuss how to obtain $\mathcal{R}_{F_1}(U\rho U^\dag)$ in experiments. For qubit states, $\mathcal{R}_{F_1}$ of each individual state reduces to the norm of its off-diagonal elements~\cite{Designolle21,Piani16}, i.e., $\mathcal{R}_{F_1}(U|\phi_i\rangle\langle \phi_i| U^\dag)=2|\langle0| U|\phi_i\rangle\langle \phi_i|  U^\dag|1\rangle|$. In the following, we denote $V_i:=\mathcal{R}_{F_1}(U|\phi_i\rangle\langle \phi_i| U^\dag)$ for simplicity. Since all the states $\{|\phi_i\}_{i=1}^5$ are in the X-Z plane of Fig.~\ref{fig1}, the off-diagonal elements of these states are always real. Therefore, one has
\begin{equation}\label{eq2}
V_i=2|\langle \phi_i| U^\dag|1\rangle\langle0| U|\phi_i\rangle|=|\langle \phi_i| U^\dag\sigma_x U|\phi_i\rangle|,
\end{equation}
where one can directly obtain each $V_i$ in experiments by  measuring the interference visibility of  Mach–Zehnder interferometer (see \cite{SM}).
In order to obtain $\mathcal{R}_1$ based on Eq. (\ref{eq1}), one needs to find the minimization over all possible unitary operators $U$. It is worth noting that the minimal bases for the set $\vec{\varrho_1}$ and $\vec{\varrho_2}$ must be in the X-Z plane  of Bloch sphere. Thus, we can restrict ourselves to using only real unitary operators for $U$, i.e., we only consider
\begin{equation}\label{eq3}
  U_\alpha=\left(\begin{matrix}
  \cos2\alpha&\sin2\alpha\\
  \sin2\alpha&-\cos2\alpha\end{matrix}\right)
\end{equation}
to implement the unitary operation of changing the basis.

For the set of $\vec{\varrho_1}$, the two states in the set are $|\phi_{1}\rangle$ and $|\phi_{2}\rangle$. Calculating $V_i$ for each of these states in different basis, we find $V_1=|\sin 4\alpha|$ and $V_2=|\cos 4\alpha|$.
Based on Eq. (\ref{eq1}), one can obtain
\begin{equation}\label{eq5}
\mathcal{R}_1(\vec{\varrho_1})=\min_\alpha\frac{V_1+V_2}{2}=\frac{1}{2}.
\end{equation}
For  the set of $\vec{\varrho_2}$, the three states selected are $|\phi_{3}\rangle$, $|\phi_{4}\rangle$, and $|\phi_{5}\rangle$. We calculate the corresponding $V_i$ for these three states in different basis and get $V_{3}=\frac{1}{3}|2\sqrt{2} \cos 4\alpha +\sin 4\alpha|$, $V_{4}=|\sin4\alpha|$ and $V_{5}=\frac{1}{3}|2\sqrt{2} \cos 4\alpha -\sin 4\alpha|$.
Similar to Eq. (\ref{eq5}), one can obtain
\begin{equation}\label{eq7}
\mathcal{R}_1(\vec{\varrho_2})=\min_\alpha\frac{V_3+V_4+V_5}{3}=\frac{10\sqrt{2}}{27}\approx 0.524.
\end{equation}

\section{Experimental setup}
First, we denote the horizontally polarized state $|H\rangle$ as  $|0\rangle$, and vertically polarized state $|V\rangle$ as $|1\rangle$. The experimental setup, shown in Fig.~\ref{fig2}, consists of three main parts.

The first part (I) is the heralded single-photon source \cite{Kim0325,Shi0426}, which provides the required photon pairs for the experiment. We use a periodically poled $\text{KTiOPO}_{4}$ (PPKTP) crystal with dimensions of $1\times2\times20$ mm, a poling period of $10.025~\mu\text{m}$, and a working temperature controlled at 25 $^{\circ}$C. It utilizes collinear type II spontaneous parametric down-conversion (SPDC) to convert a photon with the wavelength of 405 nm into two photons with 810 nm wavelength. One photon
is collected into the single mode fiber by the fiber coupler (FC) after a long-wave pass filter (LPF), and directly measured by a single photon detector (SPD) as the idler photon (trigger), while the other enters parts (II) and (III) as the signal photon. In the following, we denote $\alpha_{k}$ to represent the angle rotating the half-wave plate (HWP) $\text{H}_{k}$ shown in Fig.~\ref{fig2}, and $\beta_{k}$ to represent the angle rotating the quarter-wave plate (QWP) $\text{Q}_{k}$.

The second part (II) is the initial state preparation and the basis selection. The wave plate group (a) consists of a QWP $\text{Q}_{1}$ and a HWP $\text{H}_{1}$, which together with the polarizing beam splitter (PBS) regulates the number of photons entering part (III), ensuring that only H-polarized photons can reach the wave plate group (b). The wave plate group (b) comprises two HWPs $\text{H}_{2}$ and $\text{H}_{3}$. After passing through $\text{H}_{2}$, our initial state becomes
\begin{equation}
    |\phi_i\rangle=\cos2\alpha_2|0\rangle+\sin2\alpha_2|1\rangle.
\end{equation}
By adjusting the angle of $\text{H}_{2}$, we can obtain any  initial pure state located in the X-Z plane of Bloch sphere. The role of $\text{H}_{3}$ is to change the basis, corresponding to $U_{\alpha}$ in Eq.~(\ref{eq3}).

The third part (III) consists of a Sagnac interferometer with a 50:50 non-polarizing beamsplitter (BS). One can experimentally obtain each $V_i$ in Eq. (\ref{eq2}) by measuring the interference visibility of this Sagnac interferometer (see  \cite{SM}), thereby obtaining the robustness of  coherence for different states in different basis.
Each path of the  Sagnac interferometer contains a set of wave plate groups (c) or (d) to implement different unitary operators $U_c$ or $U_d$, respectively.

For an initial state $|\phi_i\rangle$ entering the interferometer, it can be seen that the corresponding average number of output photons is $N(|\phi_i\rangle)=\frac{1}{2}(1+\mathrm{sin}\omega|\langle \phi_i|U_{\alpha_3}^\dagger U_{c}^{\dagger}U_{d}U_{\alpha_3}|\phi_i\rangle |)$ (see  \cite{SM}). Here, $\omega$ represents the phase difference between the two arms of the interferometer. It is evident that adjusting the angle of the phase shifter, allows us to achieve both the maximum and minimum counts of photons in the experiment. As the angle $\omega$ changes, we can obtain $N_{\mathrm{max}}(|\phi_i\rangle) = \frac{1}{2}(1+|\langle \phi_i| U_{\alpha_3}^\dag U_{c}^{\dagger}U_{d} U_{\alpha_3}|\phi_i\rangle|)$ and $N_{\mathrm{min}}(|\phi_i\rangle) = \frac{1}{2}(1-|\langle \phi_i| U_{\alpha_3}^\dag U_{c}^{\dagger}U_{d} U_{\alpha_3}|\phi_i\rangle|)$, where $U_{\alpha_3}$ is $U_{\alpha}$ in Eq. (\ref{eq3}) with $\alpha=\alpha_3$. In Fig. \ref{fig2}, we set the angles of the waveplates as $\beta_{2}=\beta_{3}=45^\circ$, $\alpha_{4}=\alpha_5=-45^\circ$ for $U_{c}$, and set  $\beta_{4}=\beta_{5}=45^\circ$, $\alpha_{6}=\alpha_{7}=90^\circ$ for $U_{d}$. Thus, the obtained unitary matrix is $U_c=-\sigma_x$ and $U_d=i \mathrm{I}$, where $\sigma_x$ is the Pauli matrix and $\mathrm{I}$ is $2\times 2$ identity matrix.
When the initial state is $|\phi_i\rangle$, the interferometer visibility (see \cite{SM})
\begin{equation}\label{eeq7}
    V_i=\frac{N_{\mathrm{max}}(|\phi_i\rangle) -N_{\mathrm{min}}(|\phi_i\rangle)}{N_{\mathrm{max}}(|\phi_i\rangle) + N_{\mathrm{min}}(|\phi_i\rangle)}=|\langle \phi_i| U_{\alpha_3}^{\dagger} \sigma_x U_{\alpha_3}|\phi_i\rangle|
\end{equation}
corresponds to Eq. (\ref{eq2}). Therefore, we can obtain each $V_i$ in Eq. (\ref{eq2}) by measuring the interferometer visibility $V_i$ in experiments.


\begin{figure}[t]
\begin{center}
\includegraphics[scale=0.35]{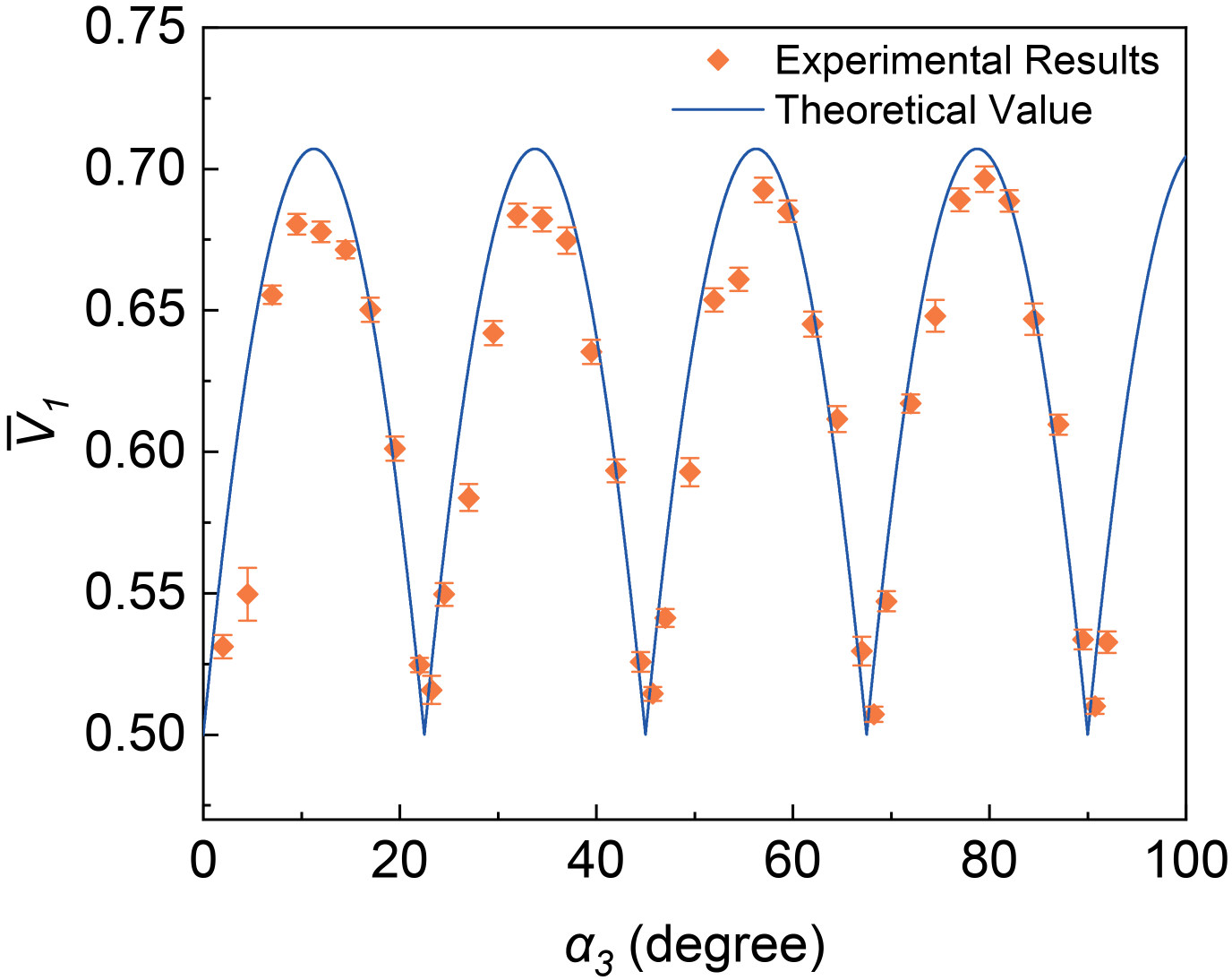}
\caption{The orange dots represent the average interference visibility  $\overline{V}_1:=(V_1+V_2)/2$ obtained from the measurements of states $|\phi_{1}\rangle$ and $|\phi_{2}\rangle$. The value range of $\alpha_{3}$ is from 2$^\circ$ to 92$^\circ$ with a step size of 2.5$^\circ$. In order to obtain the minimum point, we selected four additional angles of 23.25$^\circ$, 45.75$^\circ$, 68.25$^\circ$ and 90.75$^\circ$. A total of 41 experimental data points were obtained. The blue line represents the theoretical curve obtained by averaging the theoretical values for states $|\phi_{1}\rangle$ and $|\phi_{2}\rangle$. The experimentally obtained minimum value is $\mathcal{R}^{exp}_1(\vec{\varrho_1})=0.507 \pm 0.003$, corresponding to $\alpha_{3}$ at $68.25^\circ$.} \label{fig3}
\end{center}
\end{figure}

\section{Experimental results} 
Since we use the heralded single-photon source and do not require quantum entanglement, we ensure that the photons pass through the PPKTP crystal clockwisely with a horizontal polarization state by rotating the waveplates in (I).
In the experiment, the Sagnac interferometer visibility reached 99.4\% when both $U_c$ and $U_d$ initially set as identity matrices, indicating that our constructed optical path yields reliable data.

In the case of $n=2$, we selected two initial states $|\phi_{1}\rangle$ and $|\phi_{2}\rangle$. For the state $|\phi_{1}\rangle$  we adjusted the $\alpha_{2}$ to 0$^\circ$, and for the state $|\phi_{2}\rangle$ we adjusted it to 22.5$^\circ$. We varied the $\alpha_{3}$ from 2$^\circ$ to 92$^\circ$ with a step of 2.5$^\circ$. To ensure that the measured states are under the same basis, we changed $\alpha_{2}$ while keeping $\alpha_{3}$ constant, allowing us to obtain $V_{1}$  and $V_2$ for different states. It is important to note that the rotation angle of the phase shifter should not be too large, otherwise interference phenomena will disappear. By rotating $\text{H}_{3}$ to change the reference basis,  $V_{1}$ and $V_2$ we obtain here can be regarded as the sum of the norms of its off-diagonal elements, which represents the robustness of coherence for each individual state under the reference basis. Finally, by averaging over both states, we obtained Fig.~\ref{fig3}. The horizontal axis represents the rotation angle of $\text{H}_3$, and the vertical axis represents the average interference visibility $\overline{V}_1:=(V_1+V_2)/2$ obtained from the two states $|\phi_{1}\rangle$ and $|\phi_{2}\rangle$. It can be seen that within the range of values on the horizontal axis, the average interference visibility $\overline{V}_1$ is a periodic function. According to Eq. (\ref{eq1}), the actual points of interest in this figure are the four minimum points. The experimentally obtained minimum value is $\mathcal{R}^{exp}_1(\vec{\varrho_1})=0.507 \pm 0.003$, which is consistent with the theoretical value in Eq. (\ref{eq5}).

\begin{figure}[t]
\begin{center}
\includegraphics[scale=0.35]{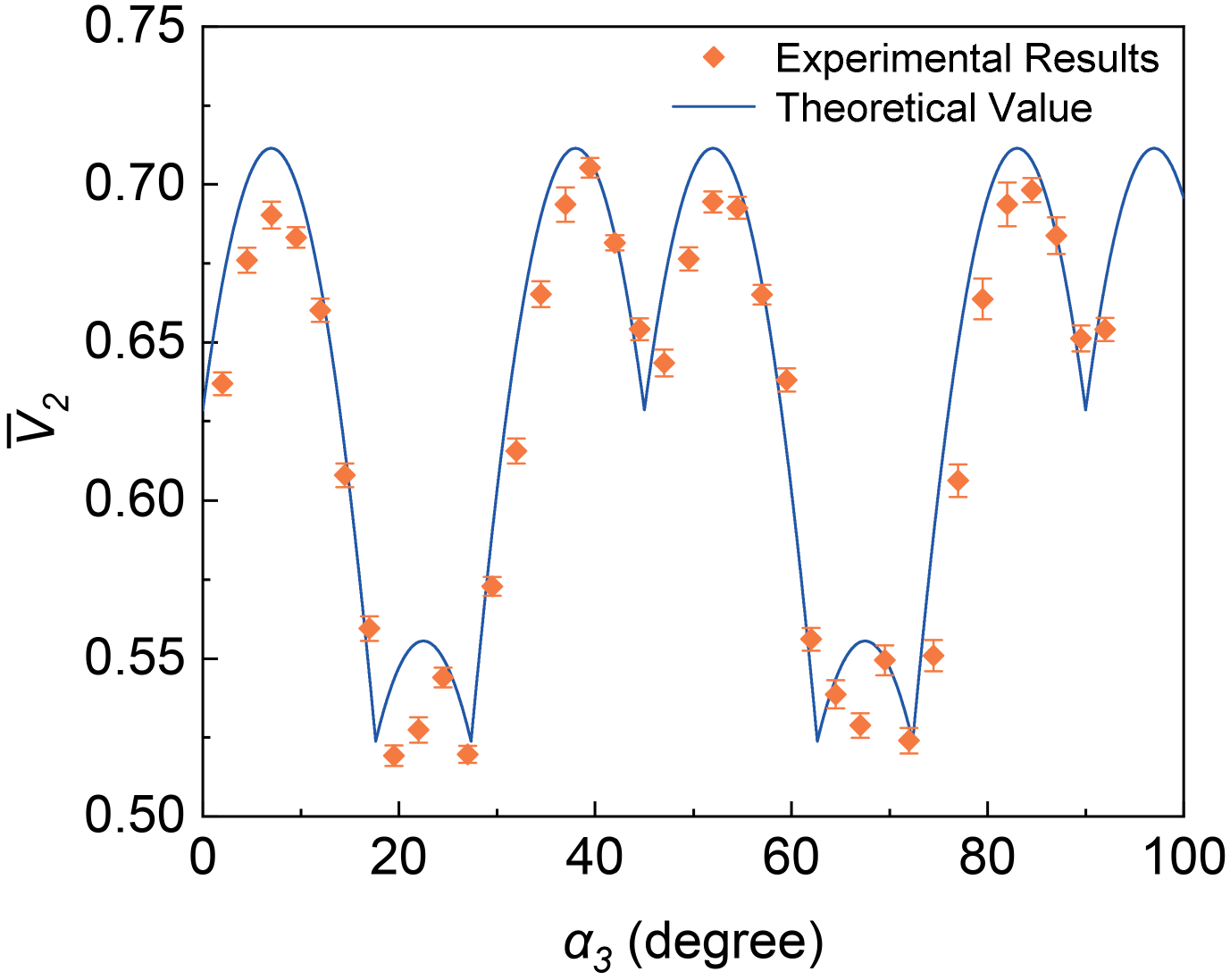}
\caption{The orange dots represent the average interference visibility $\overline{V}_2:=(V_3+V_4+V_5)/3$ obtained from the measurements of states $|\phi_{3}\rangle$ ; $|\phi_{4}\rangle$ and $|\phi_{5}\rangle$, the value range of $\alpha_{3}$ is from 2$^\circ$ to 92$^\circ$ with a step size of 2.5$^\circ$. A total of 37 experimental data points were obtained. The blue line represents the theoretical curve obtained by averaging the theoretical values for states $|\phi_{3}\rangle$ ; $|\phi_{4}\rangle$ and $|\phi_{5}\rangle$. The experimentally obtained minimum value is $\mathcal{R}^{exp}_1(\vec{\varrho_2})=0.519 \pm 0.003$, corresponding to $\alpha_{3}$ at $19.5^\circ$.} \label{fig4}
\end{center}
\end{figure}

For the case of $n=3$, we selected three initial states $|\phi_{3}\rangle$, $|\phi_{4}\rangle$, and $|\phi_{5}\rangle$. Based on Eq.~(\ref{eq7}), the theoretical value is 0.524. We adjusted the $\alpha_{2}$ to 27.4$^\circ$ to obtain $|\phi_{3}\rangle$ and set it to 0$^\circ$ to obtain $|\phi_{4}\rangle$. For the state $|\phi_{5}\rangle$, we adjusted the $\alpha_{2}$ to -27.4$^\circ$. Similarly, for $\text{H}_{3}$, we varied the $\alpha_{3}$ from 2$^\circ$ to 92$^\circ$ with a step size of 2.5$^\circ$. Finally, we averaged the three states and obtained Fig. \ref{fig4}. The horizontal axis is the rotation angle of $\text{H}_3$ as well, consisting with that in Fig. \ref{fig3}, and the vertical axis represents the average interference visibility  $\overline{V}_2:=(V_3+V_4+V_5)/3$ obtained from the three states $|\phi_{3}\rangle$, $|\phi_{4}\rangle$, and $|\phi_{5}\rangle$. The experimentally obtained minimum value is $\mathcal{R}^{exp}_1(\vec{\varrho_2})=0.519 \pm 0.003$, which is consistent with the theoretical value in Eq. (\ref{eq7}).

\begin{figure}[t]
\begin{center}
\includegraphics[scale=0.6]{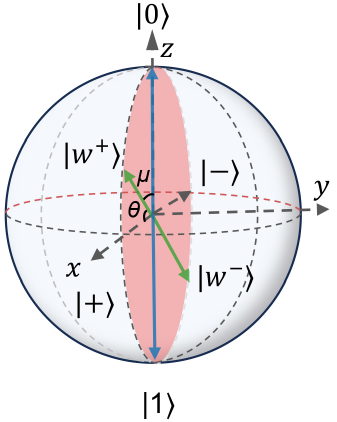}
\caption{In the non-standard BB84 protocol, Alice and Bob select the basis as either the Z basis $\{|0\rangle,|1\rangle\}$ or the W basis $\{|w^+\rangle,|w^-\rangle\}$, where the W basis is derived by rotating the ideal X basis by an angle $\theta$.} \label{fig5}
\end{center}
\end{figure}

\section{Application in BB84 protocol}
Let us now consider the application of set coherence in BB84 protocol. Suppose that we have two sets of orthogonal and complete bases, i.e., $\{|\psi(\mu_1,\varphi_1)\rangle, |\psi(\mu_1,\varphi_1)\rangle_{\perp}\}$ and $\{|\psi(\mu_2,\varphi_2)\rangle, |\psi(\mu_2,\varphi_2)\rangle_{\perp}\}$, which are defined as $|\psi(\mu_i,\varphi_i)\rangle\nonumber:=\cos\frac{\mu_i}{2} |0\rangle + e^{i\varphi_i } \sin\frac{\mu_i}{2} |1\rangle$, $|\psi(\mu_i,\varphi_i)\rangle_{\perp}:=\sin\frac{\mu_i}{2} |0\rangle - e^{i\varphi_i} \cos\frac{\mu_i}{2} |1\rangle$.
When $\mu_1=0$, $\varphi_1=\pi$,  $\mu_2=\pi/2$ and $\varphi_2=0$, the two sets of bases reduce to $\{|0\rangle, |1\rangle\}$ and $\{|+\rangle, |-\rangle\}$, which are just the bases used in BB84 protocol. The angle $\mu$ corresponds to the angle between the quantum state and the positive direction of the z-axis, as shown in Fig.~\ref{fig5}.

Now we show that the maximal set coherence of $\Phi:=\{|\psi(\mu_1,\varphi_1)\rangle, |\psi(\mu_1,\varphi_1)\rangle_{\perp}, |\psi(\mu_2,\varphi_2)\rangle, |\psi(\mu_2,\varphi_2)\rangle_{\perp}\}$ is 1/2, which can be achieved when we choose $\{|0\rangle,|1\rangle,|+\rangle,|-\rangle\}$.~One can calculate the robustness of coherence $\mathcal{R}_{F_1}$  for each individual state in $\Phi$, and obtain that $ \mathcal{R}_1(\Phi)=\min_U \frac{1}{4}\sum_{j=1}^{4} \mathcal{R}_{F_1}(U|\psi_j\rangle\langle\psi_j|U^\dag)$,
where $|\psi_j\rangle$ is the $j$-th satae in the set $\Phi$.
Furthermore, we can find that $ \mathcal{R}_{F_1}(|\psi(\mu_i,\varphi_i)\rangle)=\mathcal{R}_{F_1}(|\psi(\mu_i,\varphi_i)\rangle_{\perp})$.
 Therefore, $\mathcal{R}_1(\Phi)=\min_U \frac{1}{4}\sum_{j=1}^{4} \mathcal{R}_{F_1}(U|\psi_j\rangle\langle\psi_j|U^\dag)=\min_U \frac{1}{2}\sum_{i=1}^2\mathcal{R}_{F_1}(U|\psi(\mu_i,\varphi_i)\rangle\langle\psi(\mu_i,\varphi_i)|U^\dag)$,
 i.e., the set coherence of $\Phi$ is equal to the set coherence of $\{|\psi(\mu_i,\varphi_i)\rangle\}_{i=1}^2$, and the maximal set coherence of $\Phi$ is
\begin{equation}
  \max_{\{\mu_i,\varphi_i\}}\mathcal{R}_1(\Phi)=\frac{1}{2},
\end{equation}
 similar to Eq. (\ref{eq5}).
From the calculation for the set coherence of $\Phi$, we can conclude that  $\{|0\rangle, |1\rangle\}$ and $\{|+\rangle, |-\rangle\}$ is just one of the best choices of $\Phi$ for the maximal set coherence.

\begin{figure}[t]
\begin{center}
\includegraphics[scale=0.4]{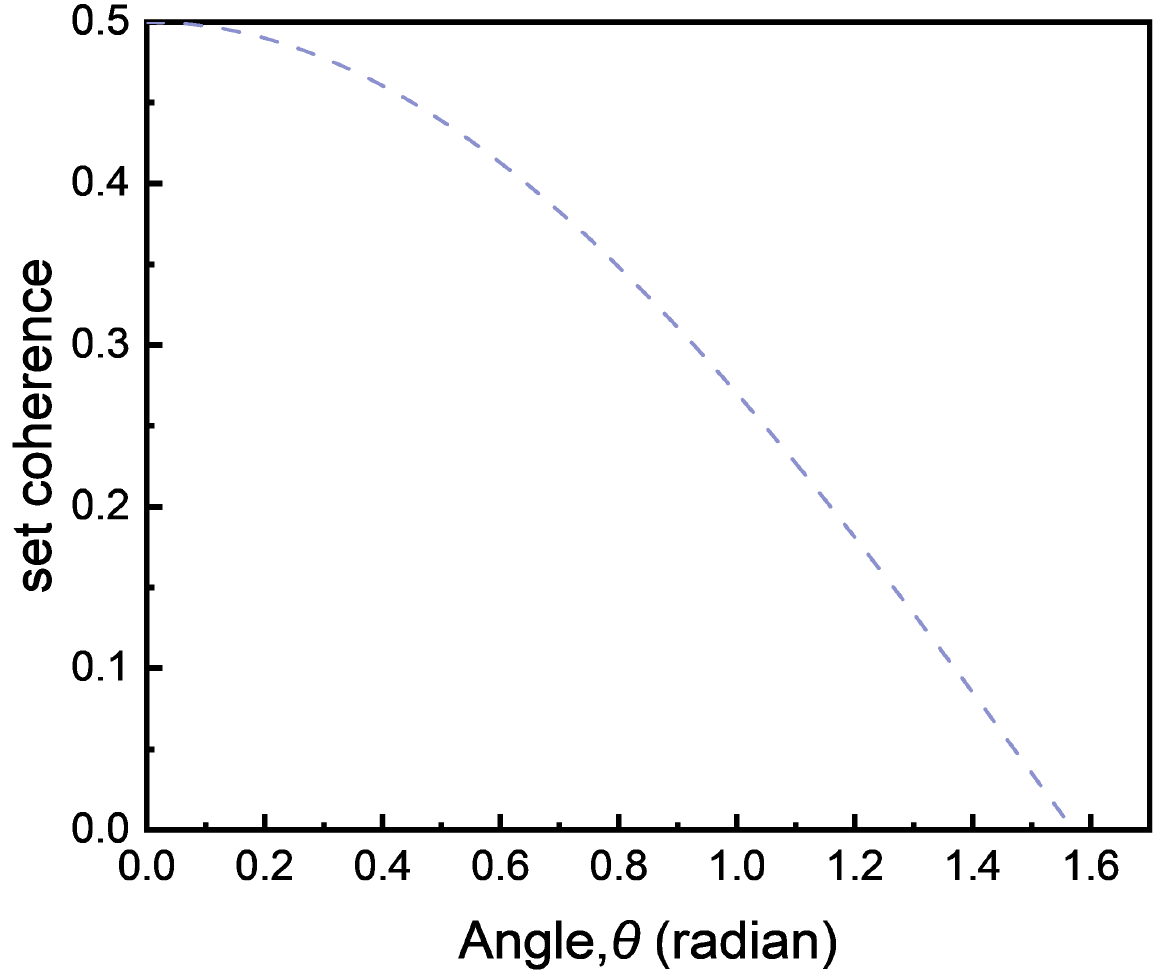}
\caption{The set coherence with the deflection angle $\theta$. As the angle $\theta$ varies from $0$ to $\pi/2$, the set coherence  is decreasing from 0.5 to 0.} \label{fig6}
\end{center}
\end{figure}

As the set coherence of the chosen states changes, how does it affect the BB84 protocol? In the standard BB84 protocol, the two selected bases are the Z and X bases, namely the states $|0\rangle$, $|1\rangle$, $|+\rangle$, and $|-\rangle$. At this point, the set coherence is maximized at 1/2. Subsequently, we consider altering the four states used in the BB84 protocol. Alice and Bob no longer employ the states $|0\rangle$, $|1\rangle$, $|+\rangle$, and $|-\rangle$, but instead use the states $|0\rangle$, $|1\rangle$, $|w^+\rangle$, and $|w^-\rangle$. As shown in Fig.~\ref{fig5}, the basis W $\{|w^+\rangle,|w^-\rangle\}$ is in the xz plane, and the angle $\theta$ corresponds to the angle between $|w^+\rangle$ and the positive direction of the x axis, thus $\mu=\pi/2 - \theta$. The two orthogonal states in the W basis are $|w^+\rangle=\text{cos}(\frac{\pi}{4}-\frac{\theta}{2})|0\rangle+\text{sin}(\frac{\pi}{4}-\frac{\theta}{2})|1\rangle$, $|w^-\rangle= \text{sin}(\frac{\pi}{4}-\frac{\theta}{2})|0\rangle-\text{cos}(\frac{\pi}{4}-\frac{\theta}{2})|1\rangle$.
 Calculations yield that for different angles $\theta$, the corresponding set coherence is
\begin{equation}
    \mathcal{R}_1(\Phi)=\frac{1}{2} \big|\sin(\frac{\pi}{2}-\theta)\big|.
\end{equation}
In Fig.~\ref{fig6}, the protocol progressively transforms from the standard BB84 protocol, as the angle $\theta$ varies from $0$ to $\pi/2$, the set coherence decreasing from 0.5 to 0. This process corresponds to Figure 5 in~\cite{Coles16}, as the angle $\theta$ increases, the key rate also decreases from its maximum value to 0. Therefore, we can deduce that higher set coherence corresponds to a higher key rate. Thus, we can assess the key rate of the protocol on the basis of the magnitude of the set coherence. Furthermore, the theory of set coherence can also be applied to other similar protocols, such as the selection of transmitting and receiving bases in quantum secure direct communication \cite{QSDC}.


\section{Discussion and conclusion}
From Figs. \ref{fig3} and \ref{fig4}, we can find that the average interference visibility $\overline{V}_1$ and $\overline{V}_2$ show a good agreement between experimental results and theoretical values. The values of set coherence, represented by the minimum value points in the graphs of $\overline{V}_1$ and $\overline{V}_2$, also have errors within acceptable ranges. For the case of $n=3$, due to limitations in the rotation precision of the wave plates, we made an approximation for the angle $\alpha_{2}$ corresponding to the prepared quantum states $|\phi_3\rangle$ and $|\phi_5\rangle$. However, the final experimental results still closely match the theoretical predictions.

In summary, we experimentally investigated the basis-independent set coherence for two separate sets of states using a Sagnac interferometer. The experimental values of set coherence are consistent with the theoretical values for both sets. Our findings may have applications in other quantum processing tasks, such as quantum cloning for a set of quantum states and quantum key distribution using a set of quantum states. For instance, similar to the application of set coherence in the BB84 protocol, one may discuss  the application of set coherence in other protocols of quantum key distribution.

\section*{Acknowledgments}
This work is supported by the National Natural Science Foundation of China (Grants No.~62475127 and No.~11875172), and Zhejiang Provincial Natural Science Foundation of China (Grant No. LZ25A040006).

\setcounter{equation}{0}
\setcounter{figure}{0}
\renewcommand{\theequation}{A\arabic{equation}}
\renewcommand{\thefigure}{S\arabic{figure}}

\section*{APPENDIX A: Implement theoretical operation}

In the study of robustness of coherence~\cite{Napoli16,Piani16,Wang17}, Ref.~\cite{Designolle21} puts forward its unique concept of quantum state set to study the coherence of a group of quantum states, and we verify the proposed concept experimentally.

\begin{figure}[t]
\begin{center}
\includegraphics[scale=0.5]{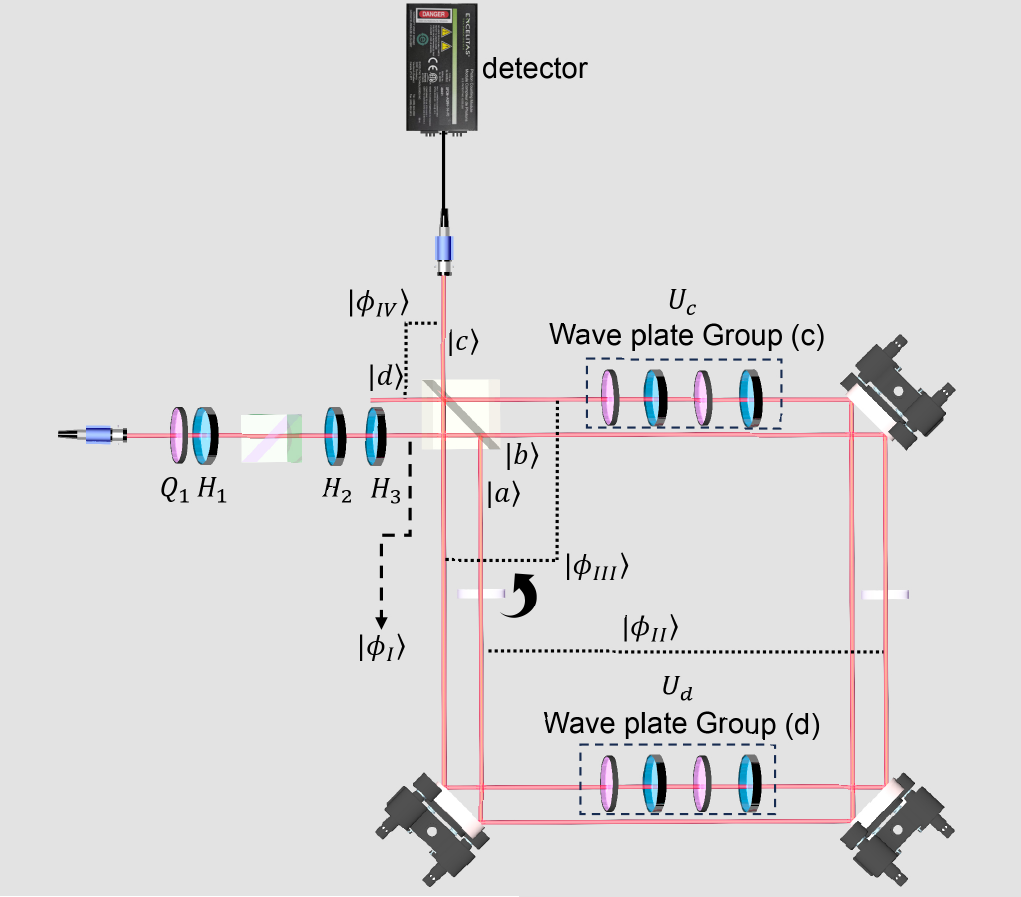}
\caption{Sagnac ring interferometer.} \label{fig21}
\end{center}
\end{figure}

\begin{table*}[t]
	\centering
	\caption{Experimental results of interference visibility for $|\phi_{1}\rangle$ under different basis}
	\label{tab:1}
	\begin{tabular}{cccc cccc cccc cccc cccc c}
		\hline\hline\noalign{\smallskip}	
		$\alpha_3$ &$2^{\circ}$ & $4.5^{\circ}$ & $7^{\circ}$ & $9.5^{\circ}$ & $12^{\circ}$ & $14.5^{\circ}$ & $17^{\circ}$ \\ &$19.5^{\circ}$ & $22^{\circ}$ & $23.25^{\circ}$ & $24.5^{\circ}$ & $27^{\circ}$ & $29.5^{\circ}$ & $32^{\circ}$ \\ &$34.5^{\circ}$ & $37^{\circ}$ & $39.5^{\circ}$ & $42^{\circ}$ & $44.5^{\circ}$ &$45.75^{\circ}$ & $47^{\circ}$   \\
		\noalign{\smallskip}\hline\noalign{\smallskip}
		$V_1$ & 0.08047~~ & 0.20605~~ & 0.44072~~&0.57999 ~~&0.70174~~ &0.81613~~& 0.88096~~\\ & 0.93452~~ & 0.98113~~& 0.99186~~&0.97114 ~~&0.92934~~&0.87905~~& 0.79720 \\ &0.67622 ~~& 0.55216~~&0.38443 ~~&0.23465~~&0.06637~~&0.03496~~&0.10180~~ \\
		\noalign{\smallskip}\hline
	\end{tabular}
	\begin{tabular}{cccc cccc cccc cccc cccc }
		\hline\hline\noalign{\smallskip}	
		$\alpha_3$ & $49.5^{\circ}$ & $52^{\circ}$ & $54.5^{\circ}$ & $57^{\circ}$ & $59.5^{\circ}$ & $62^{\circ}$ & $64.5^{\circ}$\\ & $67^{\circ}$ & $68.25^{\circ}$ & $69.5^{\circ}$ & $72^{\circ}$ & $74.5^{\circ}$ & $77^{\circ}$ & $79.5^{\circ}$\\ & $82^{\circ}$ & $84.5^{\circ}$ & $87^{\circ}$ & $89.5^{\circ}$ &$90.75^{\circ}$ & $92^{\circ}$    \\
		\noalign{\smallskip}\hline\noalign{\smallskip}
		$V_1$ & 0.23586~~ & 0.41053~~ & 0.53378~~&0.68393~~ &0.81184~~ &0.90164~~& 0.96345~~\\ & 0.98942~~& 0.99550 ~~& 0.98285~~&0.96045 ~~&0.88292~~&0.83334~~& 0.69656~~ \\& 0.56605~~ & 0.40067~~&0.27480 ~~&0.08438~~&0.02549~~&0.07954~~ \\
		\noalign{\smallskip}\hline
\end{tabular}
\end{table*}

In this section, we will introduce the calculation of interference visibility. In Fig. \ref{fig21}, we denote the state after initial state preparation through the waveplate $\text{H}_2$ as $|\phi_i\rangle$. Subsequently, this state passes through the waveplate $\text{H}_3$. The state obtained at this point is referred to as $|\phi_I\rangle$. Its specific form can be written as
\begin{eqnarray}\label{}
 |\phi_I\rangle=U_{\alpha_3} \cdot U_{\alpha_2}|H\rangle
               =U_{\alpha_3}|\phi_i\rangle.
\end{eqnarray}
The expressions $U_{\alpha_2}$ and $U_{\alpha_3}$ represent the Jones matrices corresponding to the half-wave plates $\text{H}_2$ and $\text{H}_3$, respectively. By adjusting the phase shifter within the interference loop, a relative phase $\omega$ induced by optical path difference is introduced between the two paths. When $|\phi_I\rangle$ enters the 50:50 beam splitter, its state transforms to
\begin{equation}\label{}
 |\phi_{II}\rangle=\frac{1}{\sqrt{2}}(ie^{i\omega}|\phi_I\rangle|a\rangle+|\phi_I\rangle|b\rangle).
\end{equation}
In the above equation, $|a\rangle$ and $|b\rangle$ correspond to the reflected and transmitted paths, respectively. On the reflected and transmitted paths, we place different unitary operators $U_c=-\sigma_x$ and $U_d=i \mathrm{I}$, respectively. As shown in  Fig. \ref{fig21}, they are both realized by waveplate groups consisting of half-wave plates and quarter-wave plates. After passing through the waveplate group, the state before passing through the beam splitter again can be written as
\begin{equation}\label{}
 |\phi_{III}\rangle=\frac{1}{\sqrt{2}}(ie^{i\omega}U_c|\phi_I\rangle|a\rangle+U_d|\phi_I\rangle|b\rangle).
\end{equation}
After passing through the beam splitter again, we denote the output paths as $|c\rangle$ and $|d\rangle$, respectively. The state at this moment is
\begin{eqnarray}\label{}
 |\Psi_{IV}\rangle&=&\frac12(-e^{i\omega}U_c|\phi_I\rangle+U_d|\phi_I\rangle)|c\rangle\nonumber\\
 &&+\frac12(ie^{i\omega}U_c|\phi_I\rangle+iU_d|\phi_I\rangle)|d\rangle.
\end{eqnarray}

The detector is located at port $|c\rangle$, which means we actually only detect half of the total number of photons. This can be achieved by projector $\Pi_c = |c\rangle\langle c|$ onto the entire state. At this point, the state component of measurement arm is
\begin{equation}\label{}
|\Psi_{c}\rangle=\frac{1}{2}(-e^{i\omega}U_c|\phi_I\rangle+U_d|\phi_I\rangle)|c\rangle.
\end{equation}
The intensity detected at the output port by the detector is
\begin{eqnarray}\label{}
N(|\phi_i\rangle)&=&|\langle c|\psi_{IV}\rangle|^2\nonumber\\
&=&\frac{1}{4}\Big(\langle \phi_I| U_c^{\dagger}U_c |\phi_I\rangle+\langle \phi_I| U_d^{\dagger}U_d |\phi_I\rangle\nonumber\\
&&-e^{-i\omega}\langle \phi_I| U_c^{\dagger}U_d |\phi_I\rangle
-e^{i\omega}\langle \phi_I| U_d^{\dagger}U_c |\phi_I\rangle\Big).\nonumber
\end{eqnarray}
Here, we can assume that
\begin{equation}\label{}
U_c^\dagger U_d=\sigma_x e^{-i\frac{\pi}{2}};U_d^\dagger U_c= \sigma_x e^{i\frac{\pi}{2}}.
\end{equation}
Thus, $N$ can be written as
\begin{eqnarray}\label{}
N(|\phi_i\rangle)=\frac{1}{2}(1+\langle \phi_i|U^\dagger_{\alpha_3 }\sigma_x U_{\alpha_3}|\phi_i\rangle\sin\omega).
\end{eqnarray}

Based on the number of photons detected by the detector, we can obtain the visibility $V$ of the interference. By adjusting the phase shifter angle $\omega$, we can achieve the maximum and minimum values of $N$,
\begin{eqnarray}\label{}
N_{max}(|\phi_i\rangle)&=&\frac{1}{2}(1+|\langle \phi_i|U^\dagger_{\alpha3 } \sigma_x U_{\alpha3}|\phi_i\rangle|),\\
N_{min}(|\phi_i\rangle)&=&\frac{1}{2}(1-|\langle \phi_i|U^\dagger_{\alpha3 } \sigma_x U_{\alpha3}|\phi_i\rangle|).
\end{eqnarray}
According to the definition of interference visibility
\begin{eqnarray}\label{}
V_i&=&\frac{N_{max}(|\phi_i\rangle)-N_{min}(|\phi_i\rangle)}{N_{max}(|\phi_i\rangle)+N_{min}(|\phi_i\rangle)}\nonumber\\
 &=&|\langle \phi_i| U_{\alpha_3}^{\dagger} \sigma_x U_{\alpha_3}|\phi_i\rangle|.
\end{eqnarray}

\begin{figure}[t]
\begin{center}
\includegraphics[scale=0.3]{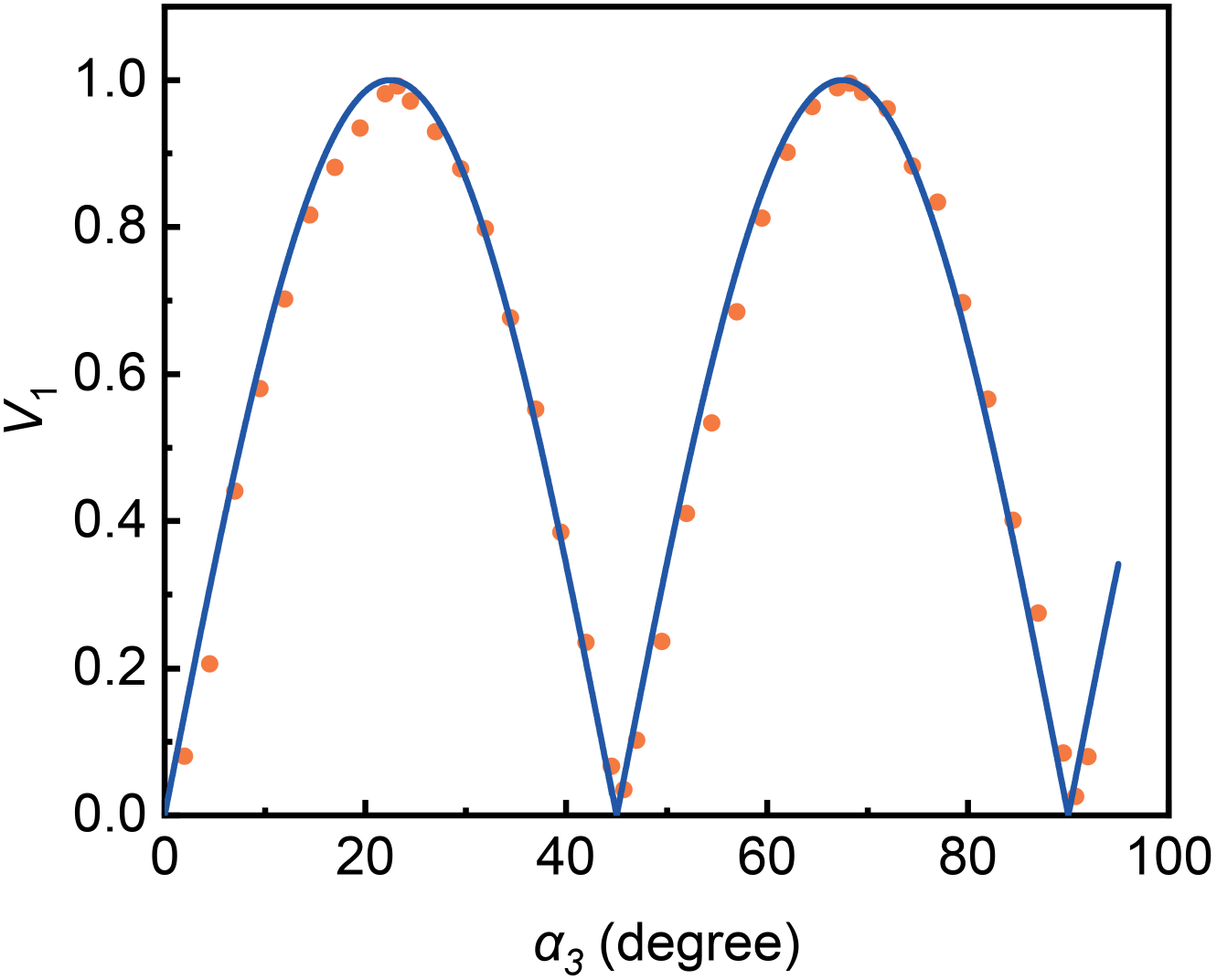}
\caption{The data points in the graph represent the experimental results of interference visibility for $|\phi_{1}\rangle$ measured under different bases, while the curves depict the theoretical predictions for $|\phi_{1}\rangle$ under different bases. The X-axis represents the deviation angle in degrees caused by $\text{H}_3$.} \label{fig22}
\end{center}
\end{figure}

\setcounter{equation}{0}
\renewcommand{\theequation}{B\arabic{equation}}
\section*{APPENDIX B: Specific calculations and experimental data}
In this section, we will illustrate the specific operation steps and show the experimental results obtained. In this experiment, we construct the target matrix by utilizing half-wave plates ($\alpha$) and quarter-wave plates ($\beta$). The symbols $\alpha$ and $\beta$ represent the angles between the fast axes of these wave plates and the horizontal polarization direction. The Jones matrices of the wave plates can be represented as
\begin{eqnarray}
    &&\text{HWP}(\alpha)=\begin{bmatrix}\cos2\alpha&\sin2\alpha\\\sin2\alpha&-\cos2\alpha\end{bmatrix},\nonumber\\\nonumber
    &&\text{QWP}(\beta)=\begin{bmatrix}\cos^2\beta+i\sin^2\beta&(1-i)\cos\beta\sin\beta\\(1-i)\cos\beta\sin\beta&\sin^2\beta+i\cos^2\beta\end{bmatrix}.
\end{eqnarray}
We also define the horizontally and vertically polarized state as
\begin{equation}
    |H\rangle=|0\rangle=\begin{bmatrix}1\\0\end{bmatrix},|V\rangle=|1\rangle=\begin{bmatrix}0\\1\end{bmatrix}.
\end{equation}
When we set $\text{H}_{2}$ to 0$^\circ$ and $\text{H}_{3}$ to $\alpha_3$ , it corresponds to the state $|\phi_{1}\rangle$. The effect of passing through these two half-wave plates on the H photons is
\begin{equation}\label{}
  |\phi^{'}_{I_{1}}\rangle=\cos2\alpha_{3}|0\rangle+\sin2\alpha_{3}|1\rangle.
\end{equation}
After the interference in the beam splitter (BS) in the optical circuit, photons pass through reflection and transmission  paths, respectively, and are subjected to different unitary matrices $U_c$ and $U_d$. After calculating the inner product of two states, the overall phase does not affect the result. Therefore, in the following content, we consider $U_c$ as $\sigma_x$ and $U_d$ as I. They then recombine at the BS, resulting in different states
\begin{eqnarray}
       |\phi^{'}_{a_1}\rangle&=&\sin2\alpha_{3}|0\rangle+\cos2\alpha_{3}|1\rangle,\\
       |\phi^{'}_{b_1}\rangle&=&\cos2\alpha_{3}|0\rangle+\sin2\alpha_{3}|1\rangle.
\end{eqnarray}
To obtain the theoretical value of the interference visibility of $|\phi_{1}\rangle$ in different bases on the X-Z plane, simply calculate and normalize the inner product of $|\phi^{'}_{a_1}\rangle$ with $|\phi^{'}_{b_1}\rangle$, resulting in
\begin{equation}\label{}
   V_1=|\sin4\alpha_3|.
\end{equation}

In the experiment, we cannot directly obtain such interference visibility. However, we can calculate the signal photon counts for larger and smaller optical paths using Eq. (7) in the main text. The results for $|\phi_{1}\rangle$ are shown in Tab. \ref{tab:1}. In order to compensate for the error in waveplate calibration, we shifted the experimental data points as a whole 2 units to the right on the X-axis. The subsequent measured data underwent the same operation. In Fig. \ref{fig22}, the theoretical curve of state $|\phi_{1}\rangle$ is shown along with the experimental data points measured.

When we set $\text{H}_{2}$ to 22.5$^\circ$ and $\text{H}_{3}$ to $\alpha_{3}$ degrees, it corresponds to the state $|\phi_{2}\rangle$. The effect of passing through these two half-wave plates on the H photons is
\begin{equation}\label{}
  |\phi^{'}_{I_2}\rangle=\frac{\cos2\alpha_{3}+\sin2\alpha_{3}}{\sqrt{2}}|0\rangle+\frac{\sin2\alpha_{3}-\cos2\alpha_{3}}{\sqrt{2}}|1\rangle.
\end{equation}
After the interference in the BS in the optical circuit, photons pass through reflection and transmission paths, respectively, and are subjected to different unitary matrices $U_c$ and $U_d$. They then recombine at the BS, resulting in different states
\begin{table*}[t]
	\centering
	\caption{Experimental results of interference visibility for $|\phi_{2}\rangle$ under different basis}
	\label{tab:2}
	\begin{tabular}{cccc cccc cccc cccc cccc }
		\hline\hline\noalign{\smallskip}	
		$\alpha_3$ &$2^{\circ}$ & $4.5^{\circ}$ & $7^{\circ}$ & $9.5^{\circ}$ & $12^{\circ}$ & $14.5^{\circ}$ & $17^{\circ}$ \\ &$19.5^{\circ}$ & $22^{\circ}$ & $23.25^{\circ}$ & $24.5^{\circ}$ & $27^{\circ}$ & $29.5^{\circ}$ & $32^{\circ}$ \\ &$34.5^{\circ}$ & $37^{\circ}$ & $39.5^{\circ}$ & $42^{\circ}$ & $44.5^{\circ}$ &$45.75^{\circ}$ & $47^{\circ}$   \\
		\noalign{\smallskip}\hline\noalign{\smallskip}
		$V_2$ & 0.98185~~ & 0.89344~~  & 0.87020 ~~&0.78090 ~~ &0.65370 ~~ &0.52673~~ & 0.41932~~ \\ & 0.26752 ~~ & 0.06823~~&0.03981 ~~&0.12817 ~~ &0.23829 ~~&0.40475~~ & 0.56998~~\\  & 0.68800  ~~& 0.79725 ~~&0.88619 ~~ &0.95177~~ &0.98521~~&0.99396~~ &0.98077~~  \\
		\noalign{\smallskip}\hline
	\end{tabular}
	\begin{tabular}{cccc cccc cccc cccc cccc }
		\hline\hline\noalign{\smallskip}	
		$\alpha_3$ & $49.5^{\circ}$ & $52^{\circ}$ & $54.5^{\circ}$ & $57^{\circ}$ & $59.5^{\circ}$ & $62^{\circ}$ & $64.5^{\circ}$\\ & $67^{\circ}$ & $68.25^{\circ}$ & $69.5^{\circ}$ & $72^{\circ}$ & $74.5^{\circ}$ & $77^{\circ}$ & $79.5^{\circ}$\\ & $82^{\circ}$ & $84.5^{\circ}$ & $87^{\circ}$ & $89.5^{\circ}$ &$90.75^{\circ}$ & $92^{\circ}$    \\
		\noalign{\smallskip}\hline\noalign{\smallskip}
		$V_2$ & 0.94964 ~~ & 0.89683  ~~& 0.78796~~ &0.70116 ~~ &0.55822 ~~ &0.38856~~ & 0.25968~~\\  & 0.06970~~& 0.01895~~ & 0.11160~~ &0.27358~~  &0.41312~~ &0.54485 ~~& 0.69619~~\\ & 0.81136~~  & 0.89300~~ &0.94426 ~~ &0.98305~~ &0.99470~~&0.98597~~  \\
		\noalign{\smallskip}\hline
\end{tabular}
\end{table*}
\begin{table*}[t]
	\centering
	\caption{Experimental results for $\Bar{V}_{1}$ under different basis}
	\label{tab:12}
	\begin{tabular}{cccc cccc cccc cccc cccc c}
		\hline\hline\noalign{\smallskip}	
		$\alpha_3$ &$2^{\circ}$ & $4.5^{\circ}$ & $7^{\circ}$ & $9.5^{\circ}$ & $12^{\circ}$ & $14.5^{\circ}$ & $17^{\circ}$ \\ &$19.5^{\circ}$ & $22^{\circ}$ & $23.25^{\circ}$ & $24.5^{\circ}$ & $27^{\circ}$ & $29.5^{\circ}$ & $32^{\circ}$ \\ &$34.5^{\circ}$ & $37^{\circ}$ & $39.5^{\circ}$ & $42^{\circ}$ & $44.5^{\circ}$ &$45.75^{\circ}$ & $47^{\circ}$   \\
		\noalign{\smallskip}\hline\noalign{\smallskip}
		$\bar{V}_1$ & 0.53116~~& 0.54975~~& 0.65546~~& 0.68044~~& 0.67772~~& 0.67143~~& 0.65014~~\\ & 0.60102~~& 0.52468~~& 0.51583~~& 0.54966~~& 0.58381~~& 0.64190~~& 0.68359 \\ & 0.68211~~& 0.67470~~& 0.63531~~& 0.59321~~& 0.52579~~& 0.51446~~& 0.54128~~ \\
        \noalign{\smallskip}\hline\noalign{\smallskip}
		error bar &0.00405~~&0.00937~~&0.00325~~&0.00362 ~~&0.00362~~&0.00297~~&0.00430~~\\ &0.00424~~ &0.00249~~&0.00497~~&0.00400~~&0.00464~~&0.00428~~&0.00409 \\ &0.00414~~&0.00470~~&0.00428~~&0.00405~~&0.00351~~&0.00246~~&0.00322~~ \\
		\noalign{\smallskip}\hline
	\end{tabular}
	\begin{tabular}{cccc cccc cccc cccc cccc }
		\hline\hline\noalign{\smallskip}	
		$\alpha_3$ & $49.5^{\circ}$ & $52^{\circ}$ & $54.5^{\circ}$ & $57^{\circ}$ & $59.5^{\circ}$ & $62^{\circ}$ & $64.5^{\circ}$\\ & $67^{\circ}$ & $68.25^{\circ}$ & $69.5^{\circ}$ & $72^{\circ}$ & $74.5^{\circ}$ & $77^{\circ}$ & $79.5^{\circ}$\\ & $82^{\circ}$ & $84.5^{\circ}$ & $87^{\circ}$ & $89.5^{\circ}$ &$90.75^{\circ}$ & $92^{\circ}$    \\
		\noalign{\smallskip}\hline\noalign{\smallskip}
		$\bar{V}_1$ & 0.59275~~ & 0.65368~~ & 0.66087~~&0.69255~~ &0.68503~~ &0.64510~~& 0.61156~~\\ & 0.52956~~& 0.50722 ~~& 0.54723~~&0.61701 ~~&0.64802~~&0.68910~~& 0.69638~~ \\& 0.68870~~ & 0.64683~~&0.60953 ~~&0.53372~~&0.51009~~&0.53275~~ \\
         \noalign{\smallskip}\hline\noalign{\smallskip}
		error bar & 0.00503~~ & 0.00411~~ & 0.00412~~&0.00438 ~~&0.00377~~ &0.00443~~& 0.00459~~\\ &0.00503~~ & 0.00267~~& 0.00362~~&0.00326 ~~&0.00559~~&0.00400~~& 0.00446 \\ &0.00383 ~~& 0.00548~~&0.00357 ~~&0.00347~~&0.00262~~&0.00376~~& \\
		\noalign{\smallskip}\hline
\end{tabular}
\end{table*}

\begin{figure}[t]
\begin{center}
\includegraphics[scale=0.3]{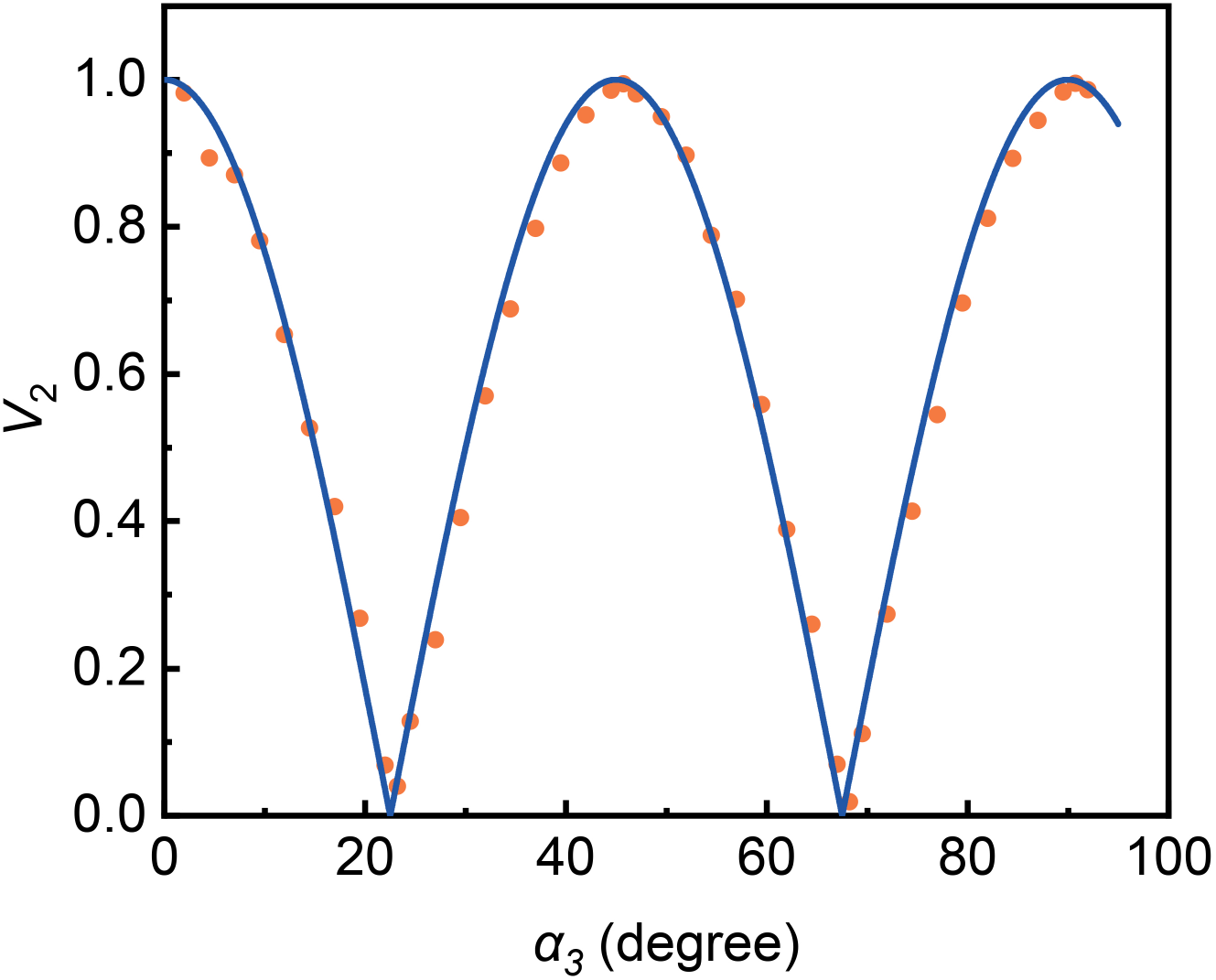}
\caption{The data points in the graph represent the experimental results of interference visibility for $|\phi_{2}\rangle$ measured under different bases, while the curves depict the theoretical predictions for $|\phi_{2}\rangle$ under different bases. The X-axis represents the deviation angle in degrees caused by $\text{H}_3$.} \label{fig23}
\end{center}
\end{figure}	

\begin{eqnarray}
       |\phi^{'}_{a_2}\rangle&=&\frac{\sin2\alpha_{3}-\cos2\alpha_{3}}{\sqrt{2}}|0\rangle+\frac{\cos2\alpha_{3}+\sin2\alpha_{3}}{\sqrt{2}}|1\rangle,\nonumber\\
       |\phi^{'}_{b_2}\rangle&=&\frac{\cos2\alpha_{3}+\sin2\alpha_{3}}{\sqrt{2}}|0\rangle+\frac{\sin2\alpha_{3}-\cos2\alpha_{3}}{\sqrt{2}}|1\rangle.\nonumber
\end{eqnarray}
To obtain the theoretical value of the interference visibility of $|\phi_{2}\rangle$ in different bases on the X-Z plane, simply calculate and normalize the inner product of $|\phi^{'}_{a_2}\rangle$ with $|\phi^{'}_{b_2}\rangle$, resulting in
\begin{equation}\label{}
  V_2=|\cos4\alpha_3|.
\end{equation}

\begin{figure}[t]
\begin{center}
\includegraphics[scale=0.3]{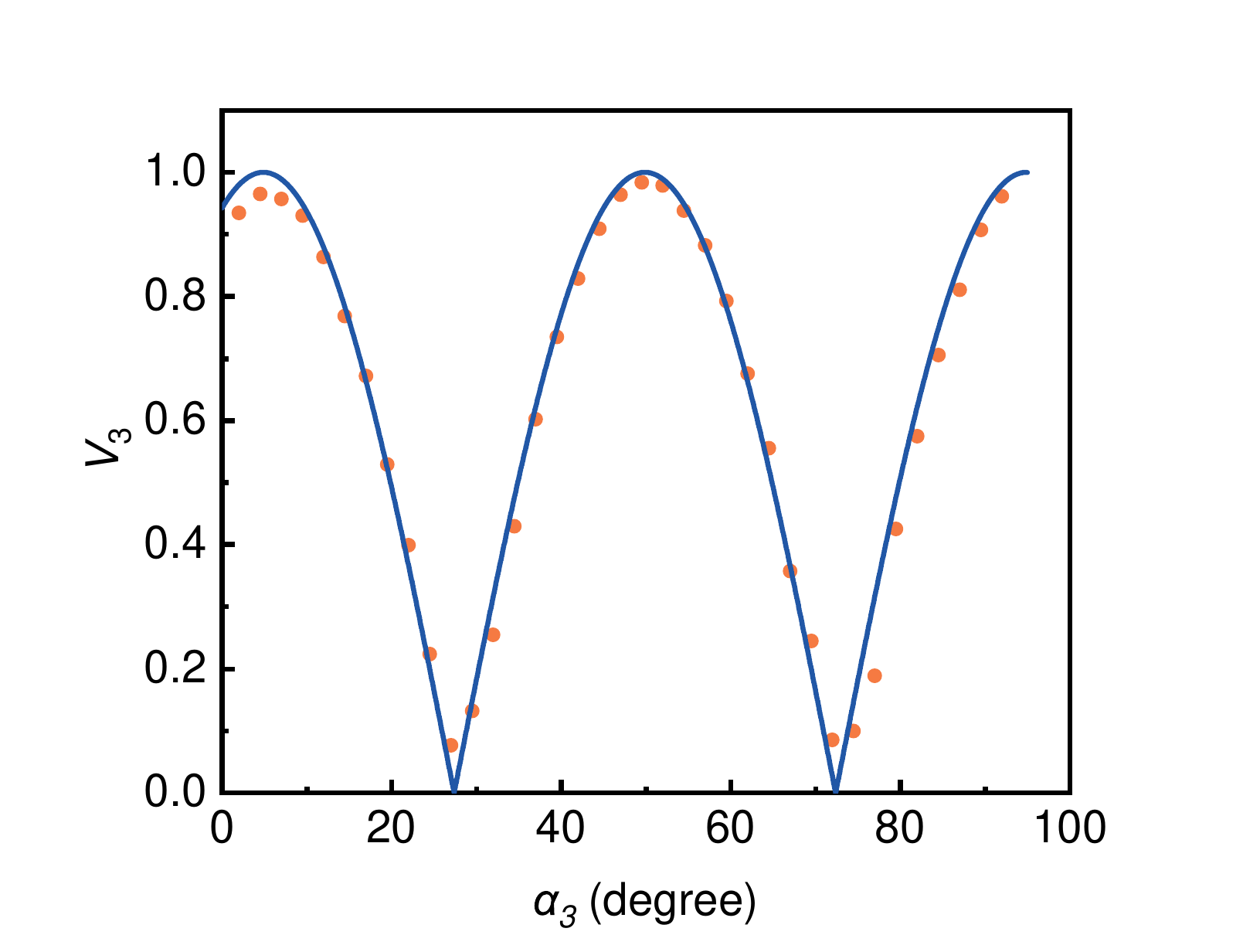}
\caption{The data points in the graph represent the experimental results of interference visibility for $|\phi_{3}\rangle$ measured under different bases, while the curves depict the theoretical predictions for $|\phi_{3}\rangle$ under different bases. The X-axis represents the deviation angle in degrees caused by $\text{H}_3$.} \label{fig24}
\end{center}
\end{figure}

\begin{figure}[t]
\begin{center}
\includegraphics[scale=0.3]{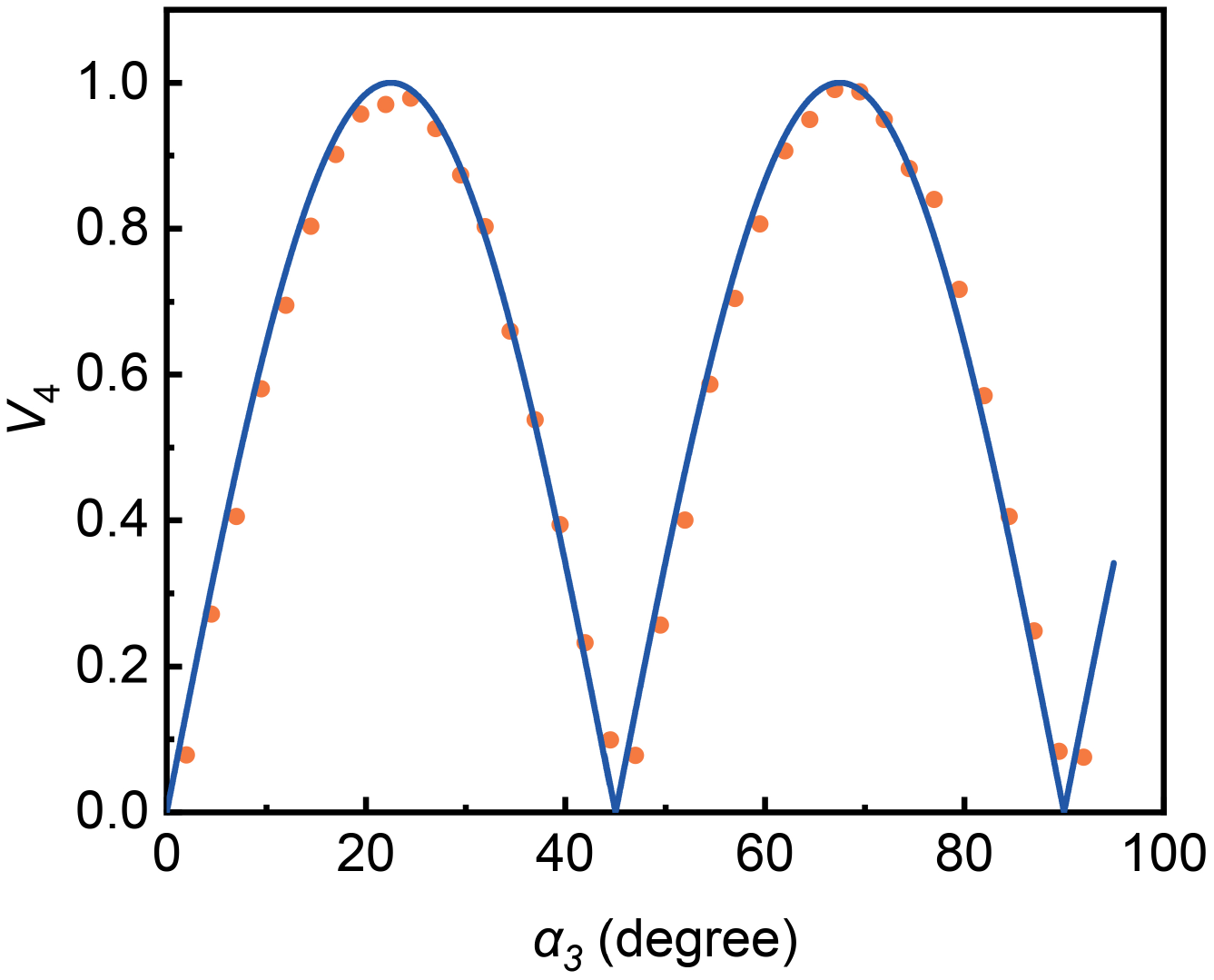}
\caption{The data points in the graph represent the experimental results of interference visibility for $|\phi_{4}\rangle$ measured under different bases, while the curves depict the theoretical predictions for $|\phi_{4}\rangle$ under different bases. The X-axis represents the deviation angle in degrees caused by $\text{H}_3$.} \label{fig25}
\end{center}
\end{figure}

The results of the experiment for $|\phi_{2}\rangle$ are shown in Tab.~\ref{tab:2}.
In Fig.~\ref{fig23}, the theoretical curve of state $|\phi_{2}\rangle$ is shown along with the experimental data points measured. Tab.~\ref{tab:12} shows the experimental results of the average visibility of interference when $n=2$.

The above two states are selected for $n=2$. For the three states selected for $n=3$, the situations are as follows. For the cases of $|\phi_{3}\rangle$ and $|\phi_{5}\rangle$, we approximate the angles of wave plates, but the following derivation process is still calculated based on the actual angle corresponding to $|\phi_{3}\rangle$ and $|\phi_{5}\rangle$.

When we set $\text{H}_{2}$ to 27.4$^\circ$ and $\text{H}_{3}$ to $\alpha_{3}$ degrees, it corresponds to the state $|\phi_{3}\rangle$. The effect of passing through these two half-wave plates on the H photons is
\begin{eqnarray}\label{}
  |\phi^{'}_{I_3}\rangle&=&\left(\frac{\cos2\alpha_{3}}{\sqrt{3}}+\sqrt{\frac23}\sin2\alpha_{3}\right)|0\rangle\nonumber\\
  &&+\left(\frac{\sin2\alpha_{3}}{\sqrt{3}}-\sqrt{\frac23}\cos2\alpha_{3}\right)|1\rangle.
\end{eqnarray}
\begin{table*}[t]
	\centering
	\caption{Experimental results of interference visibility for $|\phi_{3}\rangle$ under different basis}
	\begin{tabular}{cccc cccc cccc cccc cccc }
		\hline\hline\noalign{\smallskip}	
		$\alpha_3$ & $2^{\circ}$ & $4.5^{\circ}$ & $7^{\circ}$ & $9.5^{\circ}$& $12^{\circ}$ &$14.5^{\circ}$ & $17^{\circ}$\\ & $19.5^{\circ}$ & $22^{\circ}$ & $24.5^{\circ}$ & $27^{\circ}$& $29.5^{\circ}$ & $32^{\circ}$ & $34.5^{\circ}$\\ & $37^{\circ}$& $39.5^{\circ}$ & $42^{\circ}$ & $44.5^{\circ}$ & $47^{\circ}$   \\
		\noalign{\smallskip}\hline\noalign{\smallskip}
		$V_3$ & 0.93411~~ & 0.96505~~ &0.95676~~& 0.92987~~&0.86372~~ &0.76841~~ &0.67167~~\\& 0.52924~~ & 0.39910~~ & 0.22351~~&0.07653 ~~&0.13186~~&0.25463~~& 0.42933~~\\ & 0.60177~~ & 0.73479~~&0.82853~~ &0.90906~~&0.96371~~ \\
		\noalign{\smallskip}\hline
	\end{tabular}
	\begin{tabular}{cccc cccc cccc cccc cccc }
		\hline\hline\noalign{\smallskip}	
		$\alpha_3$ & $49.5^{\circ}$ & $52^{\circ}$ & $54.5^{\circ}$ & $57^{\circ}$& $59.5^{\circ}$ & $62^{\circ}$ & $64.5^{\circ}$\\ & $67^{\circ}$ & $69.5^{\circ}$ & $72^{\circ}$ & $74.5^{\circ}$ &$77^{\circ}$ & $79.5^{\circ}$ & $82^{\circ}$\\ & $84.5^{\circ}$ & $87^{\circ}$ & $89.5^{\circ}$ & $92^{\circ}$    \\
		\noalign{\smallskip}\hline\noalign{\smallskip}
		$V_3$ &0.98342~~& 0.97838~~ & 0.93806~~ & 0.88184~~&0.79263~~ &0.67538~~ &0.55516~~\\& 0.35710~~ & 0.24442~~ & 0.08543~~&0.09931~~ &0.18890~~&0.42517~~& 0.57438~~\\ & 0.70534~~ & 0.81025~~ &0.90724~~&0.96119~~ \\
		\noalign{\smallskip}\hline
\end{tabular}\label{tab:3}
\end{table*}

After the interference in the BS in the optical circuit, photons pass through reflection and transmission paths, respectively, and are subjected to different unitary matrices $U_c$ and $U_d$. They then recombine at the BS, resulting in different states
\begin{eqnarray}
       |\phi^{'}_{a_3}\rangle&=&\left(\frac{\sin2\alpha_{3}}{\sqrt{3}}-\sqrt{\frac23}\cos2\alpha_{3}\right)|0\rangle\nonumber\\
       &&+\left(\frac{\cos2\alpha_{3}}{\sqrt{3}}+\sqrt{\frac23}\sin2\alpha_{3}\right)|1\rangle,\\
       |\phi^{'}_{b_3}\rangle&=&\left(\frac{\cos2\alpha_{3}}{\sqrt{3}}+\sqrt{\frac23}\sin2\alpha_{3}\right)|0\rangle\nonumber\\
       &&+\left(\frac{\sin2\alpha_{3}}{\sqrt{3}}-\sqrt{\frac23}\cos2\alpha_{3}\right)|1\rangle.
\end{eqnarray}

\begin{table*}[t]
	\centering
	\caption{Experimental results of interference visibility for $|\phi_{4}\rangle$ under different basis}
	\label{tab:4}
	\begin{tabular}{cccc cccc cccc cccc cccc }
		\hline\hline\noalign{\smallskip}	
		$\alpha_3$ & $2^{\circ}$ & $4.5^{\circ}$ & $7^{\circ}$ & $9.5^{\circ}$& $12^{\circ}$ &$14.5^{\circ}$ & $17^{\circ}$\\ & $19.5^{\circ}$ & $22^{\circ}$ & $24.5^{\circ}$ & $27^{\circ}$& $29.5^{\circ}$ & $32^{\circ}$ & $34.5^{\circ}$\\ & $37^{\circ}$& $39.5^{\circ}$ & $42^{\circ}$ & $44.5^{\circ}$ & $47^{\circ}$   \\
		\noalign{\smallskip}\hline\noalign{\smallskip}
		$V_4$ & 0.07855~~ & 0.27146 ~~& 0.40540~~&0.57989~~&0.69467~~ &0.80327~~& 0.90170~~\\ & 0.95705~~ & 0.96993~~&0.97888~~ &0.93704~~&0.87349~~& 0.80239~~ & 0.65923~~\\ & 0.53756~~&0.39432~~ &0.23233~~&0.09910~~&0.07769~~ \\
		\noalign{\smallskip}\hline
	\end{tabular}
	\begin{tabular}{cccc cccc cccc cccc cccc }
		\hline\hline\noalign{\smallskip}	
		$\alpha_3$ & $49.5^{\circ}$ & $52^{\circ}$ & $54.5^{\circ}$ & $57^{\circ}$& $59.5^{\circ}$ & $62^{\circ}$ & $64.5^{\circ}$\\ & $67^{\circ}$ & $69.5^{\circ}$ & $72^{\circ}$ & $74.5^{\circ}$ &$77^{\circ}$ & $79.5^{\circ}$ & $82^{\circ}$\\ & $84.5^{\circ}$ & $87^{\circ}$ & $89.5^{\circ}$ & $92^{\circ}$    \\
		\noalign{\smallskip}\hline\noalign{\smallskip}
		$V_4$ & 0.25652~~ & 0.40033~~ & 0.58631~~&0.70391~~ &0.80591~~ &0.90663~~& 0.94947~~\\&0.99051~~ & 0.98737~~ & 0.94962~~&0.88208~~ &0.83988~~&0.71643~~& 0.57084~~\\ & 0.40554~~ & 0.24825~~&0.08329~~&0.07545~~ \\
		\noalign{\smallskip}\hline
\end{tabular}
\end{table*}
To obtain the theoretical value of the interference visibility of $|\phi_{3}\rangle$ in different bases on the X-Z plane, simply calculate and normalize the inner product of $|\phi^{'}_{a_3}\rangle$ with $|\phi^{'}_{b_3}\rangle$, resulting in
\begin{equation}\label{}
 V_{3} = \frac{1}{3}|2\sqrt{2}\cos4\alpha_3+\sin4\alpha_3|.
\end{equation}

The results of the experiment for $|\phi_{3}\rangle$ are shown in Tab. \ref{tab:3}.
In Fig. \ref{fig24}, the theoretical curve of state $|\phi_{3}\rangle$ is shown along with the experimental data points measured.

The theoretical values obtained for $|\phi_{4}\rangle$ are consistent with those in $|\phi_{1}\rangle$, and there is no need to elaborate further. The experimental data points obtained from $|\phi_{4}\rangle$ are provided in Tab. \ref{tab:4}, and the experimental points align with the theoretical curve as shown in Fig. \ref{fig25}.

\begin{figure}[h]
\begin{center}
\includegraphics[scale=0.3]{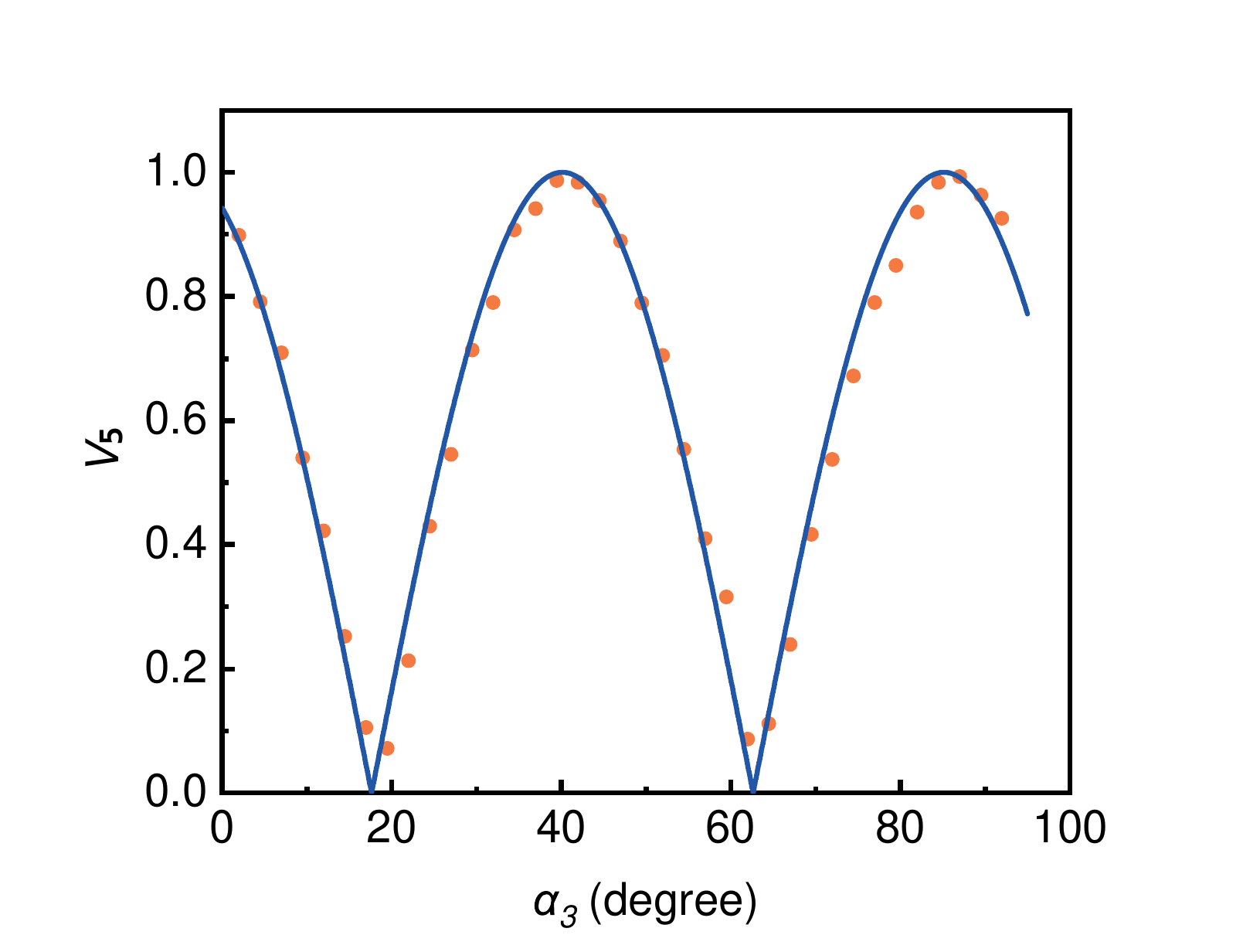}
\caption{The data points in the graph represent the experimental results of interference visibility for $|\phi_{5}\rangle$ measured under different bases, while the curves depict the theoretical predictions for $|\phi_{5}\rangle$ under different bases. The X-axis represents the deviation angle in degrees caused by $\text{H}_3$.} \label{fig26}
\end{center}
\end{figure}

\begin{table*}[h]
	\centering
	\caption{Experimental results of interference visibility for $|\phi_{5}\rangle$ under different basis}
	\label{tab:5}
	\begin{tabular}{cccc cccc cccc cccc cccc }
		\hline\hline\noalign{\smallskip}	
		$\alpha_3$ & $2^{\circ}$ & $4.5^{\circ}$ & $7^{\circ}$ & $9.5^{\circ}$& $12^{\circ}$ &$14.5^{\circ}$ & $17^{\circ}$\\ & $19.5^{\circ}$ & $22^{\circ}$ & $24.5^{\circ}$ & $27^{\circ}$& $29.5^{\circ}$ & $32^{\circ}$ & $34.5^{\circ}$\\ & $37^{\circ}$& $39.5^{\circ}$ & $42^{\circ}$ & $44.5^{\circ}$ & $47^{\circ}$   \\
		\noalign{\smallskip}\hline\noalign{\smallskip}
		$V_5$ & 0.89802~~ & 0.79141~~ & 0.70876~~&0.53989~~ &0.42216~~ &0.25198~~& 0.10512~~\\ & 0.07149~~ & 0.21318~~&0.42969~~ &0.54535~~&0.71343~~& 0.78980~~ & 0.90717~~\\ & 0.94156~~&0.98686 ~~&0.98377~~&0.95436~~&0.88902~~ \\
		\noalign{\smallskip}\hline
	\end{tabular}
	\begin{tabular}{cccc cccc cccc cccc cccc }
		\hline\hline\noalign{\smallskip}	
		$\alpha_3$ & $49.5^{\circ}$ & $52^{\circ}$ & $54.5^{\circ}$ & $57^{\circ}$& $59.5^{\circ}$ & $62^{\circ}$ & $64.5^{\circ}$\\ & $67^{\circ}$ & $69.5^{\circ}$ & $72^{\circ}$ & $74.5^{\circ}$ &$77^{\circ}$ & $79.5^{\circ}$ & $82^{\circ}$\\ & $84.5^{\circ}$ & $87^{\circ}$ & $89.5^{\circ}$ & $92^{\circ}$    \\
		\noalign{\smallskip}\hline\noalign{\smallskip}
		$V_5$ & 0.78946~~& 0.70472~~ & 0.55351~~&0.40965~~ &0.31592~~ &0.08631~~& 0.11145~~\\ & 0.23879~~ & 0.41658~~&0.53697~~ &0.67148~~&0.79011~~& 0.84967~~ & 0.93588~~ \\& 0.98389~~&0.99284~~ &0.96306~~&0.92544~~ \\
		\noalign{\smallskip}\hline
\end{tabular}
\end{table*}
\begin{table*}[t]
	\centering
	\caption{Experimental results for $\Bar{V}_{2}$ under different basis}
	\label{tab:345}
	\begin{tabular}{cccc cccc cccc cccc cccc c}
		\hline\hline\noalign{\smallskip}	
		$\alpha_3$ & $2^{\circ}$ & $4.5^{\circ}$ & $7^{\circ}$ & $9.5^{\circ}$& $12^{\circ}$ &$14.5^{\circ}$ & $17^{\circ}$\\ & $19.5^{\circ}$ & $22^{\circ}$ & $24.5^{\circ}$ & $27^{\circ}$& $29.5^{\circ}$ & $32^{\circ}$ & $34.5^{\circ}$\\ & $37^{\circ}$& $39.5^{\circ}$ & $42^{\circ}$ & $44.5^{\circ}$ & $47^{\circ}$   \\
		\noalign{\smallskip}\hline\noalign{\smallskip}
		$\bar{V}_2$ & 0.63689~~ & 0.67597~~ & 0.69031~~&0.68322 ~~&0.66018~~ &0.60789~~&0.55950~~\\ & 0.51926~~ &0.52740~~&0.54403~~&0.51964 ~~&0.57293~~&0.61561~~& 0.66524 \\ &0.69363 ~~& 0.70532~~&0.68154 ~~&0.65417~~&0.64347~~ \\
        \noalign{\smallskip}\hline\noalign{\smallskip}
		error bar & 0.00362~~ &0.00392~~ &0.00423~~&0.00330 ~~&0.00365~~ &0.00379~~& 0.00396~~\\ & 0.00321~~ & 0.00400~~& 0.00314~~&0.00266 ~~&0.00297~~&0.00395~~& 0.00408 \\ &0.00540 ~~& 0.00308~~&0.00242 ~~&0.00345~~&0.00426~~ \\
		\noalign{\smallskip}\hline
	\end{tabular}
	\begin{tabular}{cccc cccc cccc cccc cccc }
		\hline\hline\noalign{\smallskip}	
		$\alpha_3$ & $49.5^{\circ}$ & $52^{\circ}$ & $54.5^{\circ}$ & $57^{\circ}$& $59.5^{\circ}$ & $62^{\circ}$ & $64.5^{\circ}$\\ & $67^{\circ}$ & $69.5^{\circ}$ & $72^{\circ}$ & $74.5^{\circ}$ &$77^{\circ}$ & $79.5^{\circ}$ & $82^{\circ}$\\ & $84.5^{\circ}$ & $87^{\circ}$ & $89.5^{\circ}$ & $92^{\circ}$    \\
		\noalign{\smallskip}\hline\noalign{\smallskip}
		$\bar{V}_2$ & 0.67647~~ & 0.69448~~ & 0.69263~~&0.66513~~ &0.63815~~ &0.55611~~& 0.53869~~\\ & 0.52880~~& 0.54946 ~~& 0.52401~~&0.55096 ~~&0.6063~~&0.66376~~&0.6937~~ \\& 0.69826~~ & 0.68378~~&0.6512 ~~&0.65403~~ \\
         \noalign{\smallskip}\hline\noalign{\smallskip}
		error bar & 0.00371~~ & 0.00335~~ &0.00345~~&0.00311 ~~&0.00368~~ &0.00360~~& 0.00448~~\\ &0.00395~~ & 0.00478~~& 0.00400~~&0.00492 ~~&0.00508~~&0.00645~~& 0.00691 \\ &0.00384 ~~& 0.00578~~&0.00409 ~~&0.00369~~& \\
		\noalign{\smallskip}\hline
\end{tabular}
\end{table*}

When we set $\text{H}_{2}$ to -27.4$^\circ$ and $\text{H}_{3}$ to $\alpha_{3}$ degrees, it corresponds to the state $|\phi_{5}\rangle$. The effect of passing through these two half-wave plates on the H photons is
\begin{eqnarray}\label{}
  |\phi^{'}_{I_{5}}\rangle&=&\left(\frac{\cos2\alpha_{3}}{\sqrt{3}}-\sqrt{\frac23}\sin2\alpha_{3}\right)|0\rangle\nonumber\\
  &&+\left(\sqrt{\frac{2}{3}}\cos{2 \alpha_{3}}+\frac{\sin{2\alpha_{3}}}{\sqrt{3}}\right)|1\rangle.
\end{eqnarray}
After the interference in the BS in the optical circuit, photons pass through reflection and transmission paths, respectively, and are subjected to different unitary matrices $U_c$ and $U_d$. They then recombine at the BS, resulting in different states
\begin{eqnarray}
       |\phi^{'}_{a_5}\rangle&=&\left(\sqrt{\frac{2}{3}}\cos{2\alpha_{3}}+\frac{\sin{2\alpha_{3}}}{\sqrt{3}}\right)|0\rangle\nonumber\\
       &&+\left(\frac{\cos2\alpha_{3}}{\sqrt{3}}-\sqrt{\frac23}\sin2\alpha_{3}\right)|1\rangle,\nonumber\\
       |\phi^{'}_{b_5}\rangle&=&\left(\frac{\cos2\alpha_{3}}{\sqrt{3}}-\sqrt{\frac23}\sin2\alpha_{3}\right)|0\rangle\nonumber\\
       &&+\left(\sqrt{\frac{2}{3}}\cos{2\alpha_{3}}+\frac{\sin{2\alpha_{3}}}{\sqrt{3}}\right)|1\rangle.\nonumber
\end{eqnarray}
To obtain the theoretical value of the interference visibility of $|\phi_{5}\rangle$ in different bases on the X-Z plane, simply calculate and normalize the inner product of $|\phi^{'}_{a_5}\rangle$ with $|\phi^{'}_{b_5}\rangle$, resulting in
\begin{equation}\label{}
  V_{5} = \frac{1}{3}|2\sqrt{2}\cos4\alpha_3-\sin4\alpha_3|.
\end{equation}

The results of the experiment for $|\phi_{5}\rangle$ are shown in Tab.\ref{tab:5}.
In Fig. \ref{fig26}, the theoretical curve of state $|\phi_{5}\rangle$ is shown along with the experimental data points measured. Tab.\ref{tab:345} shows the experimental results of the average visibility of interference when $n=3$.

\end{document}